\documentclass[a4paper,fleqn]{cas-dc}

\usepackage[numbers]{natbib}

\def\tsc#1{\csdef{#1}{\textsc{\lowercase{#1}}\xspace}}
\tsc{WGM}
\tsc{QE}

\usepackage{url}
\usepackage{xurl}            
\usepackage{booktabs}        
\usepackage{xcolor}
\usepackage{xspace}
\usepackage{tabularx}
\usepackage{array}
\usepackage{adjustbox}
\usepackage{microtype}
\usepackage{placeins}
\usepackage{caption}
\usepackage{tikz}
\usetikzlibrary{positioning,arrows.meta,shapes.geometric}
\tikzset{
  arrow/.style={-{Latex[length=2.5mm]}, thick, draw=black!70}
}

\newcommand{\MyBox}[1]{%
  \begin{center}
  \fbox{\begin{minipage}{0.94\columnwidth}
  #1
  \end{minipage}}
  \end{center}
}

\newcommand{\totalRedditPosts}{54\xspace}

\newcommand{\nRetrievedRaw}{1383}
\newcommand{\nExcludedFirstPassRaw}{1189}
\newcommand{\nAfterFirstPassRaw}{194}
\newcommand{\nAIToolsRaw}{95}
\newcommand{\nNonCSRaw}{20}
\newcommand{\nUncertainRaw}{25}
\newcommand{\nIncludedRaw}{54}
\newcommand{\nExcludedRaw}{1329}
\newcommand{\nSubreddits}{5}
\newcommand{\nSubredditsWord}{five}
\newcommand{\nQueries}{18}
\newcommand{\nQueriesWord}{eighteen}
\newcommand{\nCombinations}{90}
\newcommand{\nCombinationsWord}{ninety}
\newcommand{\nRetrieved}{1,383}
\newcommand{\nExcludedFirstPass}{1,189}
\newcommand{\nAfterFirstPass}{194}
\newcommand{\nAITools}{95}
\newcommand{\nNonCS}{20}
\newcommand{\nUncertain}{25}
\newcommand{\nIncluded}{54}
\newcommand{\nExcluded}{1,329}
\newcommand{\nPreLLMPostsWord}{six}
\newcommand{\nPreLLMPostsWordCap}{Six}
\newcommand{\nPracticeCats}{7}
\newcommand{\nPracticeCatsWord}{seven}
\newcommand{\nSkillCats}{9}
\newcommand{\nSkillCatsWord}{nine}
\newcommand{\nNonTechSkillCatsWord}{four}
\newcommand{\nTechSkillCatsWord}{five}

\ifnum\numexpr\nRetrievedRaw-\nExcludedFirstPassRaw\relax=\nAfterFirstPassRaw\else
  \PackageWarningNoLine{numbers}{Retrieved - excluded in first pass != after first pass}\fi
\ifnum\numexpr\nAfterFirstPassRaw-\nAIToolsRaw-\nNonCSRaw-\nUncertainRaw\relax=\nIncludedRaw\else
  \PackageWarningNoLine{numbers}{After first pass - AI tools - non-CS - uncertain != included}\fi
\ifnum\numexpr\nRetrievedRaw-\nIncludedRaw\relax=\nExcludedRaw\else
  \PackageWarningNoLine{numbers}{Retrieved - included != excluded}\fi
\ifnum\numexpr\nSubreddits*\nQueries\relax=\nCombinations\else
  \PackageWarningNoLine{numbers}{Subreddits x queries != combinations}\fi

\begin{document}
\let\WriteBookmarks\relax
\def\textpagefraction{.001}
\def\floatpagepagefraction{0.85}
\shorttitle{Students' Research Practices and Skills in the LLM Era}

\shortauthors{E.R.S. de Freitas et~al.}

\title [mode = title]{A Grey Literature Review of Graduate Students' Research Practices and Skills in the LLM Era}

\author[1]{Enne Rebeca Silva de Freitas}[orcid=0009-0006-2244-0754]
\cormark[1]
\ead{enne.freitas@icen.ufpa.br}
\credit{Conceptualization, Methodology, Investigation, Data curation, Formal analysis, Writing - Original draft}

\author[2]{Breno Alves}[orcid=0009-0007-9364-9317]
\ead{baa3@cesar.school}
\credit{Conceptualization, Methodology, Writing - Review & editing}

\author[2]{Danilo Ribeiro}[orcid=0000-0001-7393-729X]
\ead{dmr@cesar.school}
\credit{Conceptualization, Methodology, Formal analysis, Validation, Supervision, Writing - Review & editing}

\author[1]{Gustavo Pinto}[orcid=0000-0003-4900-6369]
\ead{gpinto@ufpa.br}
\credit{Conceptualization, Methodology, Formal analysis, Validation, Supervision, Writing - Review & editing}

\affiliation[1]{organization={Federal University of Pará},
                city={Belém},
                state={Pará},
                country={Brazil}}

\affiliation[2]{organization={CESAR School},
                city={Recife},
                state={Pernambuco},
                country={Brazil}}

\cortext[cor1]{Corresponding author}

\begin{abstract}
\noindent\textbf{Context:} Graduate students in computing now use Large Language Models (LLMs) throughout their research, from literature reviews and coding to data analysis and writing. Yet there is little shared understanding of which skills competent use of these tools requires, so graduate programs struggle to teach, mentor, or assess it, and unverified outputs can propagate into the scientific record.\\
\textbf{Objective:} We investigate which LLM-based practices graduate students employ in their research and which skills they need to benefit from these practices.\\
\textbf{Method:} We conducted a grey literature review of Reddit discussions. We retrieved \nRetrieved{} posts from \nSubredditsWord{} research-oriented subreddits, screened them against explicit inclusion and exclusion criteria, and analyzed the \nIncluded{} included posts with thematic analysis, in which two authors independently coded each excerpt.\\
\textbf{Results:} We identified \nPracticeCatsWord{} categories of practices and \nSkillCatsWord{} categories of skills. Commenters locate the problem in accepting generated output without inspection rather than in using LLMs, and warn that delegating the effortful parts of a task, such as debugging, reading, or validating sources, delivers the output without building the skill the task was meant to develop. Traditional skills, such as programming and critical thinking, are what allow students to verify what an LLM produces.\\
\textbf{Conclusions:} Our categories give graduate programs a concrete basis to teach students to work \emph{with} LLMs without eroding the skills that research training should build, and they add research-specific skills, such as data sharing and scientific reproducibility, to existing AI competency frameworks.
\end{abstract}

\begin{highlights}
\item Analysis of 54 Reddit threads reveals seven LLM-based research practice categories.
\item Nine technical and non-technical skill categories support competent LLM usage.
\item Main risk is unverified acceptance of outputs rather than LLM adoption itself.
\item Delegating effortful tasks yields outputs without developing underlying research skills.
\item Traditional skills like programming and critical thinking enable LLM output validation.
\end{highlights}


\maketitle

\section{Introduction}
\label{sec:introduction}

Generative Artificial Intelligence (AI) tools such as ChatGPT, GitHub Copilot,
and Gemini have become embedded in the daily workflow of graduate
students~\citep{haldar2025generativeai, kosmyna2025brain, shah2025copilot}. Built on Large Language Models (LLMs), these tools now support not only writing tasks~\citep{connellpensky2025}, but also tasks across the empirical research lifecycle, from literature reviews and study design to coding and data analysis~\citep{gomes2026,trinkenreich2026}.

However, widespread availability does not imply competent use~\citep{annapureddy2025}. Recent evidence suggests that students frequently accept AI-generated content with limited verification, struggle to assess the validity of its outputs, and over-rely on these tools even in their own areas of expertise~\citep{gerlich2025,kosmyna2025brain}. In research training, this is a problem because a task is not only a deliverable, but also the means through which the student learns to be a researcher: a student who delegates a literature synthesis or a data analysis to an LLM may obtain the result without learning how to obtain it, and without being able to tell whether it is correct.

This is a problem with stakes for multiple actors. For students, over-reliance on generative AI has been associated with degradation of critical thinking~\citep{gerlich2025}, weakened metacognitive engagement~\citep{fan2025metacognitive}, and reduced ability to transfer skills to new tasks~\citep{bastani2024}. For universities and advisors, the absence of a shared understanding of what \emph{competent} AI use looks like makes it difficult to design coherent training, mentoring, and assessment practices~\citep{yan2024promises}. For the broader research community, researchers who produce outputs without internalizing the underlying skills threaten the reliability of the research they publish.

Recent literature has begun to address these concerns, but mostly from adjacent angles~\citep{yan2024promises}. A few studies have proposed AI literacy frameworks for higher education, identifying core competencies such as prompt engineering, ethical awareness, and critical evaluation of AI outputs~\citep{walter2024,annapureddy2025}. Others have measured the effects of generative AI on academic performance, reporting both productivity gains in
graduate writing tasks~\citep{connellpensky2025} and concerns about the gap
between performance and actual learning~\citep{yan2024promises,fan2025metacognitive}.

What remains underexplored is the specific case of \emph{graduate students
conducting empirical research}. Graduate research differs from coursework in at least three important ways: 1) it is open-ended, 2) it is methodologically demanding, and 3) the costs of unreflective AI use (e.g., hallucinated citations, biased syntheses, and unreproducible analyses) propagate directly into the scientific record. Existing AI literacy frameworks were not designed with these stakes in mind, and the competencies they propose seldom map onto the practical struggles graduate students face when integrating LLMs into, for instance, systematic reviews, code review, or empirical analyses.

Minimizing this gap requires evidence grounded in first-person accounts of graduate research, rather than top-down lists of competencies. Online communities offer this kind of evidence. On Reddit, research-focused communities such as \texttt{r/PhD} gather students, advisors, and researchers who, under pseudonyms, discuss how they use LLMs and the difficulties they face, in terms that rarely surface in formal publications or institutional surveys~\citep{proferes2021studying}.

In this paper, we report a Grey Literature Review (GLR)~\citep{garousi2019guidelines} of Reddit discussions on the use of LLMs in graduate research. We collected \nRetrieved{} posts from \nSubredditsWord{} research-oriented subreddits with \nQueriesWord{} queries and screened them in two stages against explicit inclusion and exclusion criteria, retaining \nIncluded{} posts. Following the thematic analysis procedure of \citet{cruzes2011thematic}, two authors independently assigned codes to each excerpt (a post or one of its comments), merged their codes, and grouped them into categories. We address
two research questions:

\begin{itemize}
\item \textbf{RQ1:} What are the \textit{LLM-based practices} that graduate students employ in their research?
\item \textbf{RQ2:} What are the \textit{skills} that graduate students must have to take better advantage of these LLM-based practices?
\end{itemize}

While \textbf{RQ1} grounds the inquiry in concrete behavior, asking what graduate students actually \emph{do} with LLMs in their research, in \textbf{RQ2} we move to the competencies needed to perform these practices well.

Our main findings are the following:

\begin{itemize}
\item \textbf{LLM use spans the whole research lifecycle, far more than writing.} We identified \nPracticeCatsWord{} categories of practices, ranging from code development, writing, and data analysis to testing, data documentation, and learning, indicating that LLMs have become part of the everyday infrastructure of graduate research rather than an isolated writing aid.

\item \textbf{The recurring concern is validation, not usage.} Across practices, commenters locate the problem in accepting generated output without inspection rather than in using LLMs. The consequences they report range from a bibliography in which four of ten sources did not exist to a university that now requires an oral examination at the end of the degree.

\item \textbf{Some reliance is accessibility, not dependence.} In a small number of comments, students who report dyslexia, Attention Deficit Hyperactivity Disorder (ADHD), or other disabilities describe LLMs as a means of access to research work rather than a way to avoid effort, a use that discussions of over-reliance rarely account for. Although this evidence is limited, it suggests that responses such as mandatory oral assessment may need accommodations for these students.

\item \textbf{Using LLMs well requires traditional skills, not only LLM-specific ones.} We identified \nSkillCatsWord{} categories of skills, \nNonTechSkillCatsWord{} non-technical (e.g., critical thinking and autonomous learning) and \nTechSkillCatsWord{} technical (e.g., programming and managing LLM context). In the accounts we analyzed, traditional skills, such as programming and critical thinking, are what allow students to verify what an LLM produces.

\item \textbf{Delegating the struggle delivers the output, not the skill.} Commenters repeatedly warn that delegating the effortful parts of a task, such as debugging, reading, or validating sources, lets students deliver the result without developing the skill the task was meant to build, which in turn leaves them less able to verify future LLM outputs. As one commenter put it, ``\emph{you can't outsource the struggle and still get the skill}''.
\end{itemize}

This paper makes three contributions: (i) an empirically grounded set of \nPracticeCatsWord{} categories of LLM-based practices and \nSkillCatsWord{} categories of skills in graduate research, derived bottom-up from \nIncluded{} Reddit discussions; (ii) a comparison of these categories with an existing AI competency framework~\citep{annapureddy2025}, showing research-specific skills, such as data sharing and scientific reproducibility, that it does not cover; and (iii) recommendations for graduate programs on how to teach, mentor, and assess the use of LLMs in research without compromising the skills research training should build. All queries, scripts, data, and qualitative coding materials are available in our replication package (see the Data availability statement).

\section{Related work}
\label{sec:related}

Our study lies at the intersection of three lines of work: studies of how LLMs are changing research practice, studies of how they affect learning and skill development, and frameworks that define the competencies needed to use them. We discuss each in turn, then the use of Reddit as a data source, and close by comparing our findings with this literature.

LLMs have moved from writing aids to tools used across the research process. By early 2024, an estimated 17.5\% of the sentences in computer science abstracts had been substantially modified by an LLM~\citep{liang2024mapping}, and a survey of 2,534 researchers in Denmark found that junior researchers use generative AI more than senior ones, and that researchers accept it more readily for language editing and data analysis than for study design or peer review~\citep{andersen2025genai}. Software engineering follows the same trend: LLMs now support data analysis, code generation, summarization, and other empirical research tasks~\citep{he2026,gomes2026}, and are reshaping the motivations, practices, and perceptions of the research community~\citep{trinkenreich2026}. With this adoption, responsibility moves to the researcher, since LLM outputs still require human validation~\citep{he2026,gomes2026} and threats such as non-reproducible outputs and data leakage can only be controlled by someone who knows where to look~\citep{sallou2024breaking}. What these studies leave open is what a graduate student, who is still learning to perform these tasks, needs to know to delegate them to an LLM.

This question matters because, for students, the effort that LLMs remove is often the effort through which skills are built. Students who used ChatGPT improved their writing while engaging less in self-regulated learning, an effect described as \emph{metacognitive laziness}~\citep{fan2025metacognitive}, and students who practiced with unrestricted access to generative AI performed worse once access was removed~\citep{bastani2024}. Frequent reliance on AI tools is also associated with cognitive offloading and lower critical thinking~\citep{gerlich2025}, and LLM-assisted essay writing with lower cognitive engagement~\citep{kosmyna2025brain}. Critical thinking does not disappear, however; it shifts toward verifying and integrating AI output, and knowledge workers who trust generative AI more report engaging in it less~\citep{lee2025critical}. Verification, in turn, depends on what the user already knows. Novice programmers who already knew what to write benefited from AI suggestions, whereas those who struggled finished with an illusion of competence~\citep{prather2024widening}. In research, the cost of failed verification is concrete, as ChatGPT fabricates a substantial share of the bibliographic references it generates~\citep{walters2023fabrication}.

If competent use cannot be taken for granted, it has to be taught, and several frameworks define what it requires. They converge on a core of AI literacy, prompting, critical evaluation of AI outputs, and awareness of ethical and legal implications~\citep{annapureddy2025,walter2024}, and educators have proposed bringing LLMs into the classroom together with reflection on their limitations, such as biases, plagiarism, and hallucinations~\citep{walter2024,yan2024promises}. Because these frameworks are derived top-down and target students or citizens in general, they say little about competencies that matter specifically in research, such as keeping analyses reproducible or tracing where data came from.

Deriving such competencies bottom-up requires first-person accounts of research work, which formal publications rarely contain. Grey literature fills part of this gap and is increasingly used as evidence in software engineering, even though it is not peer reviewed~\citep{garousi2019guidelines,kamei2020grey,kamei2022grey}. Among grey sources, Reddit has been used as a corpus in hundreds of studies across disciplines~\citep{proferes2021studying}, with documented guidelines on consent, anonymity, and data handling~\citep{FieslerZPGJ24} that we follow in this study (Section~\ref{sec:ethics}). Reddit studies on ChatGPT in education, however, have focused on general sentiment toward the tool~\citep{emdad2023chatgpt}, not on the practices and skills that graduate research requires.

Our findings confirm parts of this literature and extend others. As in surveys of researchers~\citep{andersen2025genai}, commenters are more willing to delegate support tasks than core research decisions; unlike those surveys, in which language editing is among the most accepted uses, code development is the practice with the most excerpts and posts in our corpus, ahead of writing, suggesting that in computing graduate research LLMs act first as programming partners. The trade-off reported in education~\citep{fan2025metacognitive,prather2024widening} reappears in research tasks such as literature synthesis, data analysis, and reviewing, including the illusion of competence, which commenters describe as ``more confidence than actual knowledge'', and the circularity that verifying an LLM requires the very skills its use may prevent from developing. Beyond confirming this trade-off, our data show how students try to manage it: they write the parts of the code that determine whether results are correct, delegate boilerplate, cross-check outputs across models, and consider delegation acceptable only when they could do the task themselves and verify the result. Compared with existing frameworks~\citep{annapureddy2025,walter2024}, our categories add research-specific skills, such as data sharing and scientific reproducibility, and a use of LLMs that frameworks and discussions of over-reliance overlook: access to research work for students with dyslexia or ADHD. These contributions come at a cost: our evidence is self-reported and comes from a single platform, whereas surveys and controlled experiments~\citep{andersen2025genai,bastani2024} offer larger samples and causal evidence. We see our categories as a basis on which such studies can build, and we compare them in detail with the competencies of \citet{annapureddy2025} in Section~\ref{sec:discussion}.

\section{Methodology}
\label{sec:methodology}

This study adopts a qualitative approach based on a GLR, following the guidelines proposed by \citet{garousi2019guidelines}. The investigation aims to identify and analyze information available from non-traditional scientific publication sources, such as institutional documents, technical reports, and content published on websites. The collected data were analyzed using Thematic Analysis, following the six phases proposed by \citet{cruzes2011thematic}, to identify, analyze, and interpret recurring patterns and themes related to the phenomenon under investigation. We refer to the resulting themes as \emph{categories}, each grouping a set of \emph{codes} assigned to excerpts.

The study follows four steps: Selection Pilot (Section~\ref{sec:selection}), Search (Section~\ref{sec:search}), Inclusion and Exclusion Criteria
(Section~\ref{sec:criteria}), and Thematic Analysis (Section~\ref{sec:synthesis}).

\subsection{Research Questions}
\label{sec:rqs}
The primary goal of this study is to investigate how the formative process of postgraduate researchers in Computing is being shaped by the increasing use of Large Language Models (LLMs), based on evidence from grey literature. Specifically, the study seeks to understand how LLMs influence academic supervision practices, the development of research competencies, the assessment of students' research, and their preparation for scientific careers. These objectives are reflected in the following research questions, which guide this study.

\MyBox{
    \textbf{RQ1} -- What are the LLM-based practices that graduate students employ in their research?\\[0.5em]
    \textbf{RQ2} -- What are the skills that graduate students must have to take better advantage of these LLM-based practices?
}

\subsection{Source Selection Pilot}
\label{sec:selection}

Before conducting the main search, we ran a pilot to assess three online discussion platforms as potential sources: Hacker News, Twitter/X, and Reddit. During the pilot, we also used Google Search only to refine the initial query strings (e.g., removing overly broad population terms and adding research-activity anchors such as \textit{research workflow}), not as a source of posts. \textbf{Hacker News} was discarded after most pilot queries returned threads focused on AI career decisions rather than research competencies.

\textbf{Twitter/X} was assessed through four access strategies: 1) Google \texttt{site:} search, 2) the internal GraphQL endpoint with guest token and browser cookies, 3) six Nitter instances\footnote{Instances that provided access to the content of Twitter, now known as X~\citep{nucleo_nitter_2024}.}, and 4) the \texttt{twscrape} library. All four failed due to platform restrictions introduced since 2023. \textbf{Reddit} was retained as the sole source after its public JSON API proved accessible without authentication and returned results that were substantively relevant to the research objectives.

Reddit also offers established communities focused on postgraduate study and research, including \texttt{r/PhD}, and a pseudonymous discussion environment in which users may discuss difficulties and gaps in their research skills. These characteristics motivated its selection as a source of first-person accounts concerning research competencies and practices.

\subsection{Conduct Search for Primary Studies}
\label{sec:search}

Our primary data were identified through a search conducted on Reddit using its public JSON search endpoint\footnote{\texttt{/r/\{subreddit\}/search.json}}. Data were collected on \textbf{April 20, 2026}, targeting five subreddits: \texttt{r/MachineLearning}, \texttt{r/academia}, \texttt{r/GradSchool}, \texttt{r/PhD}, and \texttt{r/compsci}. These communities were selected due to their relevance to postgraduate education, computing, and academic research.

Our search strategy targeted three core domains: postgraduate education, artificial intelligence, and the development of research competencies and practices. To broaden coverage, we formulated \nQueriesWord{} natural-language queries addressing these domains. The search endpoint and examples of the queries are presented in Figure~\ref{fig:redditsearch}. The complete set of \nQueriesWord{} queries is available in the replication package.

\begin{figure}
  \centering
  \fbox{%
    \begin{minipage}{0.94\linewidth}
      \scriptsize\raggedright
      \textbf{Search source:} Reddit public JSON search endpoint\\
      \textbf{Endpoint:} \texttt{/r/\{subreddit\}/search.json}

      \smallskip
      \textbf{Target communities:} \texttt{r/MachineLearning}, \texttt{r/academia}, \texttt{r/GradSchool}, \texttt{r/PhD}, \texttt{r/compsci}

      \smallskip
      \textbf{Search queries examples:}
      \begin{itemize}
        \setlength{\itemsep}{0pt}
        \setlength{\parskip}{0pt}
        \setlength{\topsep}{1pt}
        \setlength{\leftmargini}{1.1em}
        \item ``AI skills PhD students software engineering''
        \item ``generative AI research workflow doctoral students''
        \item ``LLM empirical research software engineering''
        \item ``generative AI data analysis research methods''
        \item ``PhD students AI competencies empirical study''
        \item ``graduate students LLM academic writing''
        \item ``AI overreliance PhD research skills''
        \item ``AI hallucination academic research competencies''
      \end{itemize}
      \textbf{Coverage:} \nSubreddits{} subreddits $\times$ \nQueries{} queries $=$ \nCombinations{} search combinations.
    \end{minipage}%
  }
  \caption{Reddit search strategy, including the search endpoint, target communities, and examples of search queries.}
  \label{fig:redditsearch}
\end{figure}

The \nQueriesWord{} queries were applied across the \nSubredditsWord{} selected subreddits, resulting in \nCombinationsWord{} subreddit--query combinations. After deduplication, \textbf{\nRetrieved{} unique posts} were identified for screening. The scripts, data, and queries are available in the replication package (see the Data availability statement).

The collection and selection procedure is summarized in four stages in Figure~\ref{fig:phases}. First, the search strategy comprised \nQueriesWord{} natural-language queries applied to \nSubredditsWord{} subreddits, resulting in \nCombinationsWord{} subreddit--query combinations. Second, data were collected on April 20, 2026, using Reddit's public JSON search endpoint. Third, after removing duplicates, \textbf{\nRetrieved{} unique posts} were retrieved and screened for relevance to the research objectives. Of these, \textbf{\nExcluded{} posts} were excluded, and \textbf{\nIncluded{} posts} met the inclusion criteria. Finally, the resulting corpus comprised \textbf{\nIncluded{} posts} distributed across \texttt{r/PhD} (14), \texttt{r/GradSchool} (5), \texttt{r/MachineLearning} (16), \texttt{r/academia} (13), and \texttt{r/compsci} (6), and was used for qualitative analysis. The criteria and two-stage screening procedure are detailed in Section~\ref{sec:criteria}.
\begin{figure}
  \centering
  \begin{adjustbox}{max width=\linewidth}
  \begin{tikzpicture}[node distance=0.14cm]
    \tikzset{flowbox/.style={draw=black!65, rounded corners=2pt,
      fill=blue!5, align=center, text width=7.0cm,
      inner sep=2.5pt, font=\scriptsize}}
    \node[flowbox] (p1) {\textbf{1. Search Strategy}\\
       \nSubreddits{} subreddits $\times$ \nQueries{} queries $=$ \nCombinations{} combinations};
    \node[flowbox, below=of p1] (p2) {\textbf{2. Data Collection}\\
       Reddit JSON endpoint: \texttt{/r/\{subreddit\}/search.json}\\April 20, 2026};
    \node[flowbox, below=of p2] (p3) {\textbf{3. Retrieved and Screened Posts}\\[1pt]
    \setlength{\tabcolsep}{3pt}
    \renewcommand{\arraystretch}{0.95}
    \begin{tabular}{@{}lrrr@{}}
      \textbf{Subreddit} & \textbf{Retrieved} & \textbf{Included} & \textbf{Excluded} \\
      \texttt{r/PhD} & 306 & 14 & 292 \\
      \texttt{r/GradSchool} & 296 & 5 & 291 \\
      \texttt{r/MachineLearning} & 293 & 16 & 277 \\
      \texttt{r/academia} & 262 & 13 & 249 \\
      \texttt{r/compsci} & 226 & 6 & 220 \\
      \hline
      \textbf{Total} & \textbf{\nRetrieved} & \textbf{\nIncluded} & \textbf{\nExcluded} \\
    \end{tabular}};
    \node[flowbox, below=of p3] (p6) {\textbf{4. Final Corpus}\\
      \textbf{$n=\nIncluded$ included posts}};
    \foreach \a/\b in {p1/p2,p2/p3,p3/p6} {
      \draw[arrow] (\a.south) -- (\b.north);
    }
  \end{tikzpicture}
  \end{adjustbox}
  \caption{Reddit search and selection process. Counts per subreddit refer to unique retrieved posts and the results of screening.}
  \label{fig:phases}
\end{figure}

\subsection{Inclusion and Exclusion Criteria}
\label{sec:criteria}
Inclusion and exclusion criteria were used to exclude studies that were not relevant to answering the research questions \citep{petersen2008systematic}. The \nRetrieved{} retrieved posts went through a two-stage screening. A post is \textbf{included} if it satisfies all of the following criteria:

\begin{itemize}
    \setlength{\itemsep}{2pt}
    \setlength{\parsep}{0pt}
    \item \textbf{AI-related research competencies:} The post or its comments discussed AI-related skills, competencies, or practices associated with research activities, including research workflows, academic writing, programming, data analysis, or other research tasks.
    \item \textbf{Graduate-level relevance:} The post or its comments referred to a master's or doctoral degree, graduate-level research, a thesis, academic supervision, or comparable postgraduate research experiences.
    \item \textbf{Computing relevance:} The discussion related to Computer Science or a computing activity, such as software engineering, programming, data analysis, machine learning, or the use of software tools in research, even when the author belonged to another disciplinary area.
    \item \textbf{Language:} The substantive discussion was written in English or Portuguese.
\end{itemize}

Conversely, a post is \textbf{excluded} if it matches at least one of the following criteria:

\begin{itemize}
    \setlength{\itemsep}{2pt}
    \setlength{\parsep}{0pt}
    \item \textbf{Undergraduate or K--12 education:} It focused exclusively on undergraduate or K--12 education.
    \item \textbf{AI tools:} It discussed AI tools exclusively for commercial promotion or advertising, without covering research practices, research-related competencies, or skill gaps.
    \item \textbf{Duplicates:} It was a duplicate of another retrieved post.
    \item \textbf{No content:} It contained only a link or image without sufficient substantive text for assessment.
    \item \textbf{Inaccessible:} It was no longer accessible at the time of screening.
\end{itemize}

\subsection{Data Extraction}
\label{sec:extraction}
Data extraction was performed using a spreadsheet with a single tab, organized as a data extraction matrix. Each row represents a post, while the columns record the identification number, subreddit, query, URL, status of \textit{exclusion}, \textit{include}, \textit{AI tools}, \textit{Non-Computer Science}, or \textit{uncertain}, two columns for codes related to practices (addressing RQ1), one column for skills (addressing RQ2), and two columns each for the first and second authors. Subsequently, additional columns were created for combined codes (one for practices and one for skills), followed by columns for the year of publication, categories related to practices, categories related to skills, and, finally, an identification column for each post, numbered from 1 to 54.

The extraction fields were structured according to two analytical dimensions: (1) usage practices in the era of LLMs among postgraduate students and researchers; and (2) gaps in skills and competencies in the era of LLMs among postgraduate students and researchers. Relevant excerpts from each post were analyzed to develop codes within the columns corresponding to each research question. If a post contained more than one relevant comment, each was coded under the appropriate field.

The included posts span 2014 to 2026 and concentrate in the most recent years: thirty-three were published in 2026, six in 2025, five in 2024, one in 2023, three in 2022, one in 2020, two in 2019, and one each in 2017, 2015, and 2014.
These \nPreLLMPostsWord{} posts were included because they feature discussions and accounts regarding the creation, use, and dilemmas surrounding AI, even from before the widespread release of AI tools. For instance, R09 and R10 address key concepts for personal development within the field of computing, as well as the reproduction and sharing of research data; R38 portrays the use of AI as a future promise; R43 covers machine learning (ML) and natural language processing (NLP), both subfields of AI, while R49 explores discussions on the use of AI by software engineers and students; and R53 provides a theoretical foundation for the development of an AI programming language that supports code development.

This structure enabled a systematic organization based on codes of practices and skills/competencies, consolidating the assignment of categories linked to each of the 54 included posts, while simultaneously maintaining the traceability of each post to its respective category and codes.

\subsection{Thematic Analysis}
\label{sec:synthesis}
The \totalRedditPosts included posts were analyzed following the thematic synthesis procedure proposed by \citet{cruzes2011thematic}.

\paragraph{Classification of posts}

The posts were classified by reading their body text and comments. A total of \nExcludedFirstPass{} posts were excluded in this first pass, mostly discussions about admissions and careers rather than skill development (e.g., ``\emph{Software engineer trying to contribute to ML research or publish independently -- advice?}'' and ``\emph{Shifting from `student as first author' to `mentor as last author' -- when?}''), leaving \nAfterFirstPass{} posts.
From these \nAfterFirstPass{} posts, we created further classifications to narrow down our analysis, categorizing them into: \textit{include}, \textit{AI tools}, \textit{Non-Computer Science}, or \textit{uncertain}. This stage set aside \nAITools{} posts that presented or discussed AI tools without addressing research practices or skills, \nNonCS{} posts unrelated to Computer Science, and \nUncertain{} posts classified as uncertain because the available text did not allow a definitive decision, resulting in a final corpus of \totalRedditPosts posts. We excluded the uncertain posts to keep the corpus restricted to posts whose relevance we could justify.

A total of \nUncertain{} posts, initially categorized as \textit{uncertain}, were excluded following a re-evaluation, as they did not address our RQs and failed to meet the inclusion criteria. For instance, while these posts mention research, ML, or computing, they consist of personal opinions regarding doctoral applications, discussions on corporate research lacking scientific rigor (involving LLMs), or the promotion of book clubs focused on computing. Although these posts touch upon topics relevant to computing or the use of AI, they are not essential to this research.

\paragraph{Qualitative coding of excerpts}
Our unit of analysis is the \emph{excerpt}: an original post or one of its comments. Two authors independently assigned codes to each excerpt, taking into account the title and body of the post and its comments. Because a single thread often reports more than one practice or skill, an excerpt could receive multiple codes. The two authors then met to merge their code sets, resolving overlaps by discussion and consolidating codes that named the same underlying behavior under a single label.
The supplementary material contains the qualitative coding schemes developed by the two authors, organized into columns for practices and skills, plus additional columns that present a merge of these codes.
We also used GPT-6, a language model from OpenAI, to assist in categorizing the publications according to the codes agreed on by the authors and compared its decisions with our own.

For instance, based on the established codes, the model suggested the category \emph{Risk, Plagiarism, and Ethical Misconduct}. From this, we developed the categories \emph{Critical Thinking and AI Evaluation} and \emph{Ethical and Responsible Use of AI}. However, following further analysis, we refined these into \emph{Critical thinking skills} and \emph{Ethics and responsibility skills} to emphasize that these are, in fact, skills. Regarding practices, the model suggested, for example, the category \emph{AI Literacy and LLM}, which we subsequently adapted into the category \emph{AI-assisted learning}.

\paragraph{Grouping codes into categories}
The merged codes were then grouped into higher-level categories along two dimensions: LLM-based practices, which answer RQ1, and skills, which answer RQ2. For instance, the codes \emph{facing references by AI}, \emph{grammatical review}, and \emph{independence and capability without AI} were grouped into the category \emph{AI-assisted learning}. Similarly, the codes \emph{develop coding practice}, \emph{dependency free learning}, \emph{preserve struggle as a learning method}, \emph{Practice writing and reviewing outputs}, \emph{Learn the fundamentals}, and \emph{Use AI for support tasks and not research} were grouped into the category \emph{Autonomous learning skills}. Finally, we split the skill categories into non-technical and technical skills, adapting the skill category classification proposed by Rabelo et al.~\citep{rabelo2022nontechnical}, Annapureddy et al.~\citep{annapureddy2025}, and Ribeiro et al.~\citep{ribeiro2026research}. This resulted in \nPracticeCatsWord{} categories of practices and \nSkillCatsWord{} categories of skills, \nNonTechSkillCatsWord{} non-technical and \nTechSkillCatsWord{} technical.

Because a single thread can contribute many excerpts, the number of excerpts in a category reflects both how often the category is discussed and how long the threads that discuss it are. We therefore report, for each category, both the number of excerpts and the number of distinct posts (Tables~\ref{tab:PracticesCode} and~\ref{tab:SkillsCode}), and we use these numbers to describe the corpus, not to estimate prevalence among graduate students.

Each excerpt was assigned to categories based on its codes; specifically, there are categories for practices and categories for skills. For instance, R27 includes practice-related codes such as \emph{supervising, delegating, and optimizing AI-generated code} and \emph{disclosing the use of AI in publications}, all grouped into the category \emph{AI-assisted code development}. Conversely, among the skill categories, it includes codes such as \emph{developing and understanding outputs}, \emph{learning without reliance/crutches}, and \emph{knowledge of institutional policy}, which encompass three categories: \emph{Autonomous learning skills}, \emph{Ethics and responsibility skills}, and \emph{Programming skills}.

\begin{center}
\includegraphics[width=0.95\columnwidth]{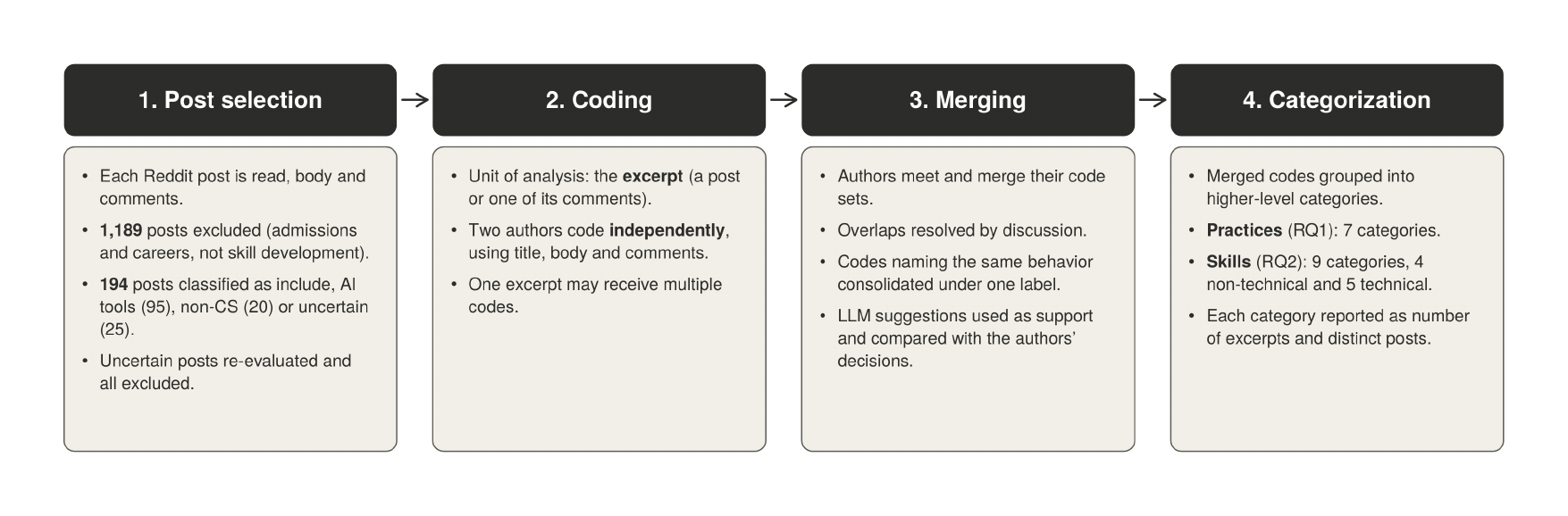}
\captionof{figure}{Overview of the three phases of the study: search, selection, and thematic analysis.}
\label{fig:Phases}
\end{center}

Figure~\ref{fig:Phases} summarizes the three phases of the study. In the search phase, we collected \nRetrieved{} posts from Reddit. In the selection phase, we first classified posts as \textit{include} or \textit{exclude} (excluding \nExcludedFirstPass{} posts), and then classified the remaining \nAfterFirstPass{} posts as \textit{include}, \textit{AI tools}, \textit{Non-Computer Science}, or \textit{uncertain}, which resulted in \nIncluded{} included posts. Finally, in the thematic analysis phase, we qualitatively coded the excerpts with respect to RQ1 and RQ2 and grouped the codes into categories.

\subsection{Ethical Considerations}
\label{sec:ethics}

We analyzed only publicly available posts and comments, collected through Reddit's public interface, and did not interact with any user. Following the recommendations of \citet{FieslerZPGJ24}, we do not report usernames, we refer to posts by identifiers (R01--R54), and we avoid linking to comments that disclose health conditions or disabilities.

\FloatBarrier
\section{RQ1: The LLM-based practices}
\label{sec:rq1}

Table~\ref{tab:PracticesCode} lists the \nPracticeCatsWord{} categories of practices we identified, with the codes consolidated in each one. We discuss them below, from the most to the least frequent.

\begin{table*}[!ht]
  \centering
  \footnotesize
  \caption{Categories of LLM-based practices (RQ1). \textit{Excerpts/Posts} indicates the number of coded excerpts (comments or posts) and the number of distinct posts (threads) they come from, respectively.}
  \label{tab:PracticesCode}
  \begin{tabularx}{\textwidth}{@{}>{\raggedright\arraybackslash}p{3cm} c >{\raggedright\arraybackslash}X@{}}
    \toprule
    \textbf{Category (\nPracticeCats)} & \textbf{Excerpts/Posts} & \textbf{Consolidated codes} \\
    \midrule
    AI-assisted code development &
    83/20 &
    Full or partial coding using LLMs; debugging and refactoring code. Automatic code generation and masking of AI-generated text; delegation of coding to AI agents. \\
    \addlinespace
    AI-assisted learning &
    35/13 &
    Use of LLMs for formulating ideas and concepts, and for learning. Use of LLMs in learning; critique of the excessive use of AI; using AI to explore adjacent fields; using AI to overcome writer's block; using AI to learn concepts from reference lists; AI-generated references; using NotebookLM for literature reviews; LLMs as search engines; questioning AI and asking for evidence; breaking tasks down into small units. \\
    \addlinespace
    AI-assisted writing and text correction &
    33/17 &
    Use of LLMs for writing, reading, and revising academic text. Restructuring, correction, and humanization of AI-generated writing; AI-generated text detection; mass submission and review; speech-to-text for capturing ideas. \\
    \addlinespace
    AI-assisted data analysis &
    10/7 &
    Use of LLMs to support data analysis; sharing of research code and data; AI-driven metadata extraction; data privacy concerns; prototyping of research code; as well as pipeline tracking and documentation. \\
    \addlinespace
    AI-assisted testing and automation &
    9/5 &
    Use of LLMs for testing, unit test generation, and AI-supervised automation; AI-assisted code writing and refactoring; code refactoring; avoiding the use of dysfunctional code. \\
    \addlinespace
    AI-assisted accessibility &
    5/2 &
    Use of AI to support writing for accessibility, dyslexia, and ADHD. \\
    \addlinespace
    Data documentation for AI systems &
    5/3 &
    Difficulty tracking data used in models; inaccurate data pipelines; lack of research reproducibility; using research repositories and documenting before coding. \\
    \bottomrule
  \end{tabularx}
\end{table*}

\paragraph{AI-assisted code development}
In the midst of AI-driven development, students rely on LLMs for coding assistance, such as writing, reviewing, and debugging code~\href{https://www.reddit.com/r/PhD/comments/1m906sy/comment/n53kwei/?utm_source=share&utm_medium=web3x&utm_name=web3xcss&utm_term=1&utm_content=share_button}{\emph{(R21)}}. 
However, one commenter observes that some students merely copy and paste AI-generated outputs, describing
\href{https://www.reddit.com/r/academia/comments/1oz70h8/comment/npmqw74/?utm_source=share&utm_medium=web3x&utm_name=web3xcss&utm_term=1&utm_content=share_button}{``\emph{graduate students who say they can ``\emph{code}'' but really can only copy and paste code and replace certain words without having a full understanding}''~(R02)}. In this account, generated code replaces, rather than supports, the student's comprehension of the problem.
As a result, the reported over-reliance on AI reveals difficulties in understanding the problem or in analyzing the correctness, efficiency, and complexity of the code. On the other hand, another student notes
\href{https://www.reddit.com/r/PhD/comments/1l7fcfu/comment/mwwhoh6/?utm_source=share&utm_medium=web3x&utm_name=web3xcss&utm_term=1&utm_content=share_button}{``\emph{I try not to copy and paste too much so I fully understand what the code is doing}''~(R27)} and \href{https://www.reddit.com/r/PhD/comments/1l7fcfu/comment/mx1sugq/?utm_source=share&utm_medium=web3x&utm_name=web3xcss&utm_term=1&utm_content=share_button}{``\emph{Use it to refine my code}''~(R27)}; both examples illustrate students who seek to understand the solutions LLMs propose for problems, including suggested corrections and new structures. Respondent R47 demonstrates a more proactive approach to understanding their workflow with LLMs:
\href{https://www.reddit.com/r/MachineLearning/comments/1s9gu8i/comment/odq36bg/?utm_source=share&utm_medium=web3x&utm_name=web3xcss&utm_term=1&utm_content=share_button}{``\emph{[...] helping me understand a code base in greater depth when starting on a new project}''~(R47)}.
These accounts suggest that commenters see AI-generated code as something to be understood and evaluated, not only accepted, which requires enough background to recognize redundant functions, duplicate code, or unnecessary structures.

To avoid programming errors, some users report employing multistage strategies when interacting with LLMs. Beyond immediate code generation, these approaches rely on initial validation and the iterative refinement of intent, as well as pipelines involving multiple models.
For example, some users prioritize core logic before optimizing the architecture, first providing context to the tool
\href{https://www.reddit.com/r/MachineLearning/comments/1sdmn97/comment/oep0bua/?utm_source=share&utm_medium=web3x&utm_name=web3xcss&utm_term=1&utm_content=share_button}{``\emph{[...] I write (ugly) code of what I want, specially evaluations, loss, etc that are very critical. Then ask agents to make the code more modular, extendable}''~(R33)}. In this workflow, the researcher keeps the parts that determine whether results are correct and delegates the parts that determine how the code is organized.
Others use cross-validation pipelines involving different models to detect hallucinations and refine output quality:
\href{https://www.reddit.com/r/PhD/comments/1kdstvd/comment/mqdfes4/?utm_source=share&utm_medium=web3x&utm_name=web3xcss&utm_term=1&utm_content=share_button}{``\emph{What I do is get code from GPT, refine it though deepseek, then ask for suggestions from GPT again}''~(R29)}. 
Other comments emphasize productivity. One commenter in R29 warns that
\href{https://www.reddit.com/r/PhD/comments/1kdstvd/comment/mqe62kq/?utm_source=share&utm_medium=web3x&utm_name=web3xcss&utm_term=1&utm_content=share_button}{``\emph{In a few years, you will be out of the game if you can’t use AI to boost your productivity}''}, and another, in R27, states that \href{https://www.reddit.com/r/PhD/comments/1l7fcfu/comment/mwytumy/?utm_source=share&utm_medium=web3x&utm_name=web3xcss&utm_term=1&utm_content=share_button}{``\emph{the main benefits are increased coding productivity, debugging, and faster coding}''}. In these comments, the pressure to adapt is framed in terms of output, and learning is not mentioned.

Another point is how AI assistants speed up coding tasks that students could perform on their own:
\href{https://www.reddit.com/r/PhD/comments/1l7fcfu/using_copilot_while_coding_feeling_guilty/?utm_source=share&utm_medium=web3x&utm_name=web3xcss&utm_term=1&utm_content=share_button}{``\emph{I definitely could write the code myself, but Copilot speeds things up a lot}''~(R27)}. In the same thread, another student treats programming as a means to an end:
\href{https://www.reddit.com/r/PhD/comments/1l7fcfu/comment/mww9ose/?utm_source=share&utm_medium=web3x&utm_name=web3xcss&utm_term=1&utm_content=share_button}{``\emph{I view coding as a means to an end for my research so if a tool helps speed up the repetitive parts and leaves me to do the higher level thinking part of the work, I don’t see it as a huge problem}''~(R27)}. Commenters who delegate in this way restrict the AI to repetitive work, such as
\href{https://www.reddit.com/r/compsci/comments/1kc9zyz/comment/mq5tj6h/?utm_source=share&utm_medium=web3x&utm_name=web3xcss&utm_term=1&utm_content=share_button}{``\emph{[...] time-wasting, no-brainer parts of my work}''~(R48)}, and still check the output; as one of them puts it,
\href{https://www.reddit.com/r/academia/comments/1rc6bt0/comment/o6w6cnh/?utm_source=share&utm_medium=web3x&utm_name=web3xcss&utm_term=1&utm_content=share_button}{``\emph{you pick random, usually critical parts of their output and check if it makes sense}''~(R03)}. Delegation is thus described as acceptable when two conditions hold: the student could do the task without the tool, and the student verifies what the tool produces.

\paragraph{AI-assisted learning}
The use of LLMs as a learning resource is a practice adopted by teachers and students, particularly in research~\citep{yan2024promises, walter2024}.
In this context, reports highlight an accelerated pace of knowledge acquisition and research execution; for instance, one commenter emphasizes the importance of using LLMs in academia \href{https://www.reddit.com/r/academia/comments/1ngj487/comment/neea57j/?utm_source=share&utm_medium=web3x&utm_name=web3xcss&utm_term=1&utm_content=share_button}{``\emph{[...] it's a great tool for research. Now you can learn things 10x as faster as the PhDs who came before you}''~(R06)}.
This comment equates learning with speed; the time that would otherwise be spent creating, questioning, and refining one's work is precisely what such acceleration may remove.
Another reported aspect is productivity in handling academic tasks, as noted by a commenter in R06: \href{https://www.reddit.com/r/academia/comments/1ox9s93/comment/np0deop/?utm_source=share&utm_medium=web3x&utm_name=web3xcss&utm_term=1&utm_content=share_button}{``\emph{We are all navigating these new waters. I use AI a lot. It's been very helpful for my administrative responsibilities, summarizing slides, helping me with transition-slide scripts, creating assessments and rubrics.}''}.
Furthermore, accounts indicate that some advisors recommend the direct use of LLMs throughout the research process, which not every student welcomes: \href{https://www.reddit.com/r/PhD/comments/1m906sy/comment/n54hw1k/?utm_source=share&utm_medium=web3x&utm_name=web3xcss&utm_term=1&utm_content=share_button}{``\emph{I'm pretty insistent of still working most of literature reviews and implementations myself, because I want to absorb things but my advisors do like to recommend using LLMs to get more immediate results as well}''~(R21)}; see also \href{https://www.reddit.com/r/PhD/s/AwCDOurEy5}{\emph{R21}} and \href{https://www.reddit.com/r/academia/s/EgFQOQXNCG}{\emph{R06}}.

Other commenters argue that students should treat LLMs with caution, because they lack the judgment of an experienced researcher: \href{https://www.reddit.com/r/PhD/s/st2I81LArf}{``\emph{We should see AI like a very knowledgeable intern. They can do basic things but they don't know how to think yet Reframe it that way}''~(R29)}. They also warn that LLMs can generate misconceptions, so a student who relies on them as the sole source of information may learn distorted or false concepts~\href{https://www.reddit.com/r/MachineLearning/comments/1lkmkuw/comment/mztwk7j/?utm_source=share&utm_medium=web3x&utm_name=web3xcss&utm_term=1&utm_content=share_button}{\emph{(R45)}}.
This trend appears in the experiences of advisors; for instance, in R01, an advisor reports that a student submitted references generated by an LLM without checking them: \href{https://www.reddit.com/r/academia/s/XRhE1ui7hh}{``\emph{[...] Out of ten sources, four led nowhere - dead links, nonexistent books, actual scholars authoring papers they never wrote}''}.

Another strategy mentioned in the reports involves using LLMs to obtain explanations and facilitate the understanding of complex topics \href{https://www.reddit.com/r/academia/comments/1r8qgac/comment/o6bf8k1/?utm_source=share&utm_medium=web3x&utm_name=web3xcss&utm_term=1&utm_content=share_button}{``\emph{if you are instead interested in just using AI in the sense of using chat bots, you can use ChatGPT or something to get interactive explanations of concepts related to your field}''~(R13)}.
In other comments students share how they use LLMs for assistance during their study process
\href{https://www.reddit.com/r/PhD/comments/1pgkhul/comment/nst16l8/?utm_source=share&utm_medium=web3x&utm_name=web3xcss&utm_term=1&utm_content=share_button}{``\emph{I don't use AI to do all my work but rather walk me through the process so I can. Well, learn in the process too}''~(R54)}.
In these accounts, the same tool supports learning when the student uses it to understand a step they then perform, and replaces learning when it performs the step for them. Because the models can fabricate concepts, explanations obtained in this way still need to be checked against other sources, such as textbooks and course material.

\paragraph{AI-assisted writing and text correction}
Using AI to draft and revise text, notably research papers, is one of the most frequently discussed practices. Alongside productivity gains, commenters raise concerns about content that appears authentic but is not true:
\href{https://www.reddit.com/r/PhD/comments/1kdstvd/comment/mqh9hmc/?utm_source=share&utm_medium=web3x&utm_name=web3xcss&utm_term=1&utm_content=share_button}{``\emph{AI will confidently create passages that don't exist...}''~(R29)}. In R02, a professor points out another concern regarding student practices, noting that \href{https://www.reddit.com/r/academia/comments/1oz70h8/comment/npbt0tn/?utm_source=share&utm_medium=web3x&utm_name=web3xcss&utm_term=1&utm_content=share_button}{``\emph{They aren't willing to put in any effort at all and keep submitting AI ``\emph{writing}'' and plagiarizing}''}. In both comments, the concern is not that the text is generated, but that the student neither checks nor engages with it.

Other identified uses of LLMs for writing and editing include conducting research \href{https://www.reddit.com/r/PhD/comments/1m906sy/comment/n54j0e6/?utm_source=share&utm_medium=web3x&utm_name=web3xcss&utm_term=1&utm_content=share_button}{``\emph{[...] some GPTs are great at finding relevant research articles for you}''~(R21)}, as well as summarizing and translating texts:
\href{https://www.reddit.com/r/GradSchool/comments/1p8rcos/comment/nr737wl/?utm_source=share&utm_medium=web3x&utm_name=web3xcss&utm_term=1&utm_content=share_button}{``\emph{I’m not an expert but text summarization and translation are two tasks that LLMs are indeed already very good}''~(R16)}. Some also use them to obtain explanations of textual content, finding it \href{https://www.reddit.com/r/PhD/comments/1kdstvd/comment/mrv4j76/?utm_source=share&utm_medium=web3x&utm_name=web3xcss&utm_term=1&utm_content=share_button}{``\emph{[...] useful when reading papers that have more technical language or help me grasp the intuition of concepts}''~(R29)}. 
These uses benefit readers who already understand the subject; a reader unfamiliar with it has no way to notice when a summary or translation distorts the original.
Another student, who uses AI to find references, worries about being accused of using it for writing: \href{https://www.reddit.com/r/academia/s/IlMw84caE5}{``\emph{I have found myself worrying over being accused of using AI, and have since tried to ``\emph{humanize}'' my own writing}''~(R04)}. The attempt to ``\emph{humanize}'' writing reveals a new form of insecurity during the research process, where the way students express their ideas and critical thinking could work against them if it resembles outputs generated by an LLM.

Some commenters argue that graduate programs, reviewers, and students alike need to adapt:
\href{https://www.reddit.com/r/academia/comments/1o0owyq/shouldnt_we_train_more_scholars_to_review_papers/}{``\emph{[...] programs still train scholars primarily to write papers, not to review them [...] What if doctoral training included structured review practice, AI literacy, and AI-ethics for reviewing?}''~(R05)}. Another lists what the academic community will need to adjust to: \href{https://www.reddit.com/r/academia/comments/1qwed3p/comment/o4bqptb/?utm_source=share&utm_medium=web3x&utm_name=web3xcss&utm_term=1&utm_content=share_button}{``\emph{1) people using suboptimal systems to make reviews aka slop, 2) how to take advantage of these systems, 3) make sure new scholars really make the most of the reviews they decide to write}''~(R08)}. Both comments shift the focus of training from producing text to evaluating it, a skill that becomes more important as text becomes cheaper to produce.

\paragraph{AI-assisted data analysis}
Commenters describe analyzing research data with LLMs as useful but risky, with reports of \href{https://www.reddit.com/r/MachineLearning/comments/1s9gu8i/comment/odr78mj/?utm_source=share&utm_medium=web3x&utm_name=web3xcss&utm_term=1&utm_content=share_button}{``\emph{people blindly following procedures to analyze data}''~(R47)}.
Prompting saves the effort of writing plotting and analysis code, as reported in R29: \href{https://www.reddit.com/r/PhD/comments/1kdstvd/comment/mqdmhxy/?utm_source=share&utm_medium=web3x&utm_name=web3xcss&utm_term=1&utm_content=share_button}{``\emph{I can just tell ChatGPT that I want a double violin plot with points simultaneously scattered under the violins then colored on a gradient [...] And in about a minute I have roughly the correct web of multi axis matplotlib soup}''}. Another commenter in R29 relies on LLMs for analysis code they could not otherwise write:
\href{https://www.reddit.com/r/PhD/comments/1kdstvd/comment/mqid86l/?utm_source=share&utm_medium=web3x&utm_name=web3xcss&utm_term=1&utm_content=share_button}{``\emph{I personally use it quite a lot for writing code to perform data analysis and fitting or numerical simulations because I'm not an expert on writing code}''}. The first case saves time for someone who could write the code; the second makes analyses possible for someone who could not, but also leaves them less able to check the result.

Concerns about who actually performed the analysis have also led to new rules for verifying a student's knowledge:
\href{https://www.reddit.com/r/PhD/comments/1kdstvd/comment/mqmn4tf/?utm_source=share&utm_medium=web3x&utm_name=web3xcss&utm_term=1&utm_content=share_button}{``\emph{People using AI to write their theses and interpret their findings is one of the reasons my uni now requires an oral exam at the end!}''~(R29)}. This comment suggests that institutions are beginning to respond by assessing whether students can explain their analyses, not only whether they produced them.

\paragraph{AI-assisted testing and automation}
Commenters also use LLMs and agents to automate tests and other repetitive tasks, weighing productivity gains against the need to understand what is automated. In the following account, a researcher reports on an automated data extraction pipeline:
\href{https://www.reddit.com/r/academia/comments/1qwed3p/comment/o3qxrl6/?utm_source=share&utm_medium=web3x&utm_name=web3xcss&utm_term=1&utm_content=share_button}{``\emph{Was only involved in software development not using the software but the test case that was published involved nearly 1000 papers and the results were the same as here, the LLM outperformed the humans at this data extraction and of course did it in a tiny tiny tiny fraction of the time}''~(R32)}.
Delegating such tasks lets researchers spend their time on problems that require judgment, since writing standardized test scripts, boilerplate code, and basic unit tests is rarely where the research contribution lies. In R32, a commenter discussing Codex argues along the same lines that LLMs should automate tedious tasks, such as testing, so that developers can focus on finding new solutions to problems~\href{https://www.reddit.com/r/MachineLearning/comments/wiqjxv/comment/ijemodc/?utm_source=share&utm_medium=web3x&utm_name=web3xcss&utm_term=1&utm_content=share_button}{(R32)}.

In R47, this appreciation for automating repetitive tasks comes with a preference for keeping control over the core of the work: \href{https://www.reddit.com/r/MachineLearning/comments/1s9gu8i/comment/odq1dp4/?utm_source=share&utm_medium=web3x&utm_name=web3xcss&utm_term=1&utm_content=share_button}{``\emph{preferring to retain control over model development, as AI-generated code can sometimes lack the nuance and optimization needed for complex ML problems}''}.
Similarly, a comment on R44, which also reflects a desire for independence, describes how the author uses AI tools while writing tests: \href{https://www.reddit.com/r/MachineLearning/comments/1mln24c/comment/n7v9h1e/?utm_source=share&utm_medium=web3x&utm_name=web3xcss&utm_term=1&utm_content=share_button}{``\emph{Sometimes, I check my process with Claude, but the actual coding part is all me}''}. 

In R50, commenters who distrust the nuances of LLM-generated tests and prefer to write them by hand~\href{https://www.reddit.com/r/compsci/comments/1oh7bx5/comment/nlmax6a/?utm_source=share&utm_medium=web3x&utm_name=web3xcss&utm_term=1&utm_content=share_button}{(R50)} still acknowledge the pressure to adapt: \href{https://www.reddit.com/r/compsci/s/MUbeCIBOd0}{``\emph{To survive this cut throat IT job market, we must adapt to changing trends and technologies and that includes getting skilled with LLM}''~(R50)}. Others accept generated tests on pragmatic grounds: \href{https://www.reddit.com/r/PhD/comments/1pgkhul/comment/nss3f1f/?utm_source=share&utm_medium=web3x&utm_name=web3xcss&utm_term=1&utm_content=share_button}{``\emph{The fact that the AI wrote the test is of little consequence assuming it collapses a 2-hour affair into a 10-minute one}''~(R54)}. The condition in this last comment matters: a generated test suite that passes, accepted without understanding what it tests, gives an illusion of quality and may hide logical flaws. Whether students write tests or not, what they need is the ability to judge what a test checks.

\paragraph{AI-assisted accessibility}
For some students, the use of LLMs is not voluntary dependence but a means of access. In R29, one student reports having a disability and using LLMs for support, and another reports using them because they have dyslexia, a condition that makes reading and writing harder; a commenter in R07 reports the same. Two other comments in R29 come from students with ADHD or concentration difficulties: one reports losing focus while working, and the other uses an LLM to think aloud, typing out thoughts as if speaking to a friend. Although these accounts come from only two posts, they show that, for these students, LLMs act as assistive technologies. Policies that restrict LLM use or require oral examinations, discussed elsewhere in the corpus, may therefore need accommodations for them.

\paragraph{Data documentation for AI systems}
Tracking data lineage before and during model training requires organization and planning, and is a precondition for reproducible research. In R36, a user states that
\href{https://www.reddit.com/r/MachineLearning/comments/1r7usv0/d_how_do_you_track_data_lineage_in_your_ml/}{``\emph{The academic literature on reproducibility keeps pointing to data provenance as a core problem, papers can't be replicated because the exact data pipeline isn't documented}''}. 
Not every researcher sees documenting and sharing code as part of the job, as seen in R44:
\href{https://www.reddit.com/r/MachineLearning/comments/1mln24c/comment/n7shorq/?utm_source=share&utm_medium=web3x&utm_name=web3xcss&utm_term=1&utm_content=share_button}{``\emph{The currency of researchers is publications, not repos. To me, a repo, it's just code that re-creates the experiments and figures that I discussed in my paper. If the idea is important enough, somebody else will put it into production. I don't even have enough SWE skills to do that competently.}''}.

In this view, the idea is the contribution and reproducing it is someone else's job, a position at odds with open science, which seeks to make data and methods accessible and transparent. Commenters in R36 see automation as one way to lower the cost of documentation:
\href{https://www.reddit.com/r/MachineLearning/comments/1r7usv0/comment/o637d0f/?utm_source=share&utm_medium=web3x&utm_name=web3xcss&utm_term=1&utm_content=share_button}{``\emph{On the zero-config automatic tracking approach. The value proposition is strong for research and prototyping where you want lineage without ceremony. [...] I didn't know it was recording that'' problem cuts both ways}''}.
As this comment notes, automatic tracking only helps if researchers know what is being recorded and why. Another comment in R36 shows that versioning alone is not enough:
\href{https://www.reddit.com/r/MachineLearning/comments/1r7usv0/comment/o61jpi5/?utm_source=share&utm_medium=web3x&utm_name=web3xcss&utm_term=1&utm_content=share_button}{``\emph{But the thing is, we used a lot of custom code and non standard data pipelines. Now, we do have all the code used on git and all the data used backed up, though}''}. 
Even when code and data are saved, custom pipelines remain hard to reconstruct. In R37, a commenter suggests AI agents as a way to automate documentation and track which files were modified~\href{https://www.reddit.com/r/MachineLearning/s/0OKMPOSXWy}{(R37)}. In all these accounts, documenting the pipeline is a means to an end: being able to defend the research and validate its results.

Table~\ref{tab:PracticesThemes} maps the excerpts to the \nPracticeCatsWord{} categories of practices. For each category, it gives a representative quote and the posts from which the coded comments come. Posts R09, R10, R14, R15, R18, R19, R23, R24, R25, R26, R35, R38, R42, R49, R51, and R53 do not appear in any practice category, but appear in the skill categories. 

\begin{table*}[H]
  \centering
  \fontsize{6.8}{7.8}\selectfont
  \renewcommand{\arraystretch}{0.85}
  \caption{Categories of practices (RQ1), with a representative quote and the posts of the coded excerpts. A number in parentheses is the number of coded excerpts from that post.}
  \label{tab:PracticesThemes}
  \begin{tabularx}{\textwidth}{@{}>{\raggedright\arraybackslash}p{2.1cm} X >{\raggedright\arraybackslash}p{3.6cm}@{}}
    \toprule
    \textbf{Practice category (\nPracticeCats)} & \textbf{Representative citation} & \textbf{Examples} \\
    \midrule
    AI-assisted code development &
    ``many students find it easier to ask ChatGPT or any similar AI tool to write the code for them, and they may add minor modifications to the code if needed, but nobody bothers themselves with writing from scratch like before'' &
    R02\,(2), R03\,(2), R13, R21\,(2), R22, R27\,(9), R29\,(5), R31, R32\,(2), R33\,(6), R34, R40, R41\,(2), R43, R44\,(12), R45\,(3), R47\,(7), R48\,(16), R50\,(3), R54\,(6) \\
    \addlinespace[2pt]
    AI-assisted writing and text correction &
    ``I left academia because professors and universities in my country encouraged academics and PhD students to use AI to write their research papers/thesis. There were workshops and guidebooks on `how to make the best AI prompts to answer your research question' and `how to turn your thesis into a book with AI' [\ldots]'' &
    R02, R04, R05, R06\,(3), R08\,(2), R12, R16, R20, R21\,(3), R22\,(2), R29\,(10), R30, R33, R37, R45\,(2), R46, R52 \\
    \addlinespace[2pt]
    AI-assisted testing and automation &
    ``Say I want to refactor my barebones model to include tests. I'll think of the folder structure on my own, write separate tests, and think of the design choices. Sometimes, I check my process with Claude, but the actual coding part is all me.'' &
    R32\,(2), R44, R47, R48, R50\,(4) \\
    \addlinespace[2pt]
    AI-assisted learning &
    ``I use LLMs frequently as a guide and generator / sounding board for ides and concepts'' &
    R01\,(2), R06\,(4), R11, R13, R17, R21\,(3), R29\,(8), R32, R45\,(7), R47\,(3), R48\,(2), R52, R54 \\
    \addlinespace[2pt]
    AI-assisted data analysis &
    ``People using AI to write their theses and interpret their findings is one of the reasons my uni now requires an oral exam at the end!'' &
    R08, R17, R22, R28, R29\,(4), R39, R47 \\
    \addlinespace[2pt]
    AI-assisted accessibility &
    ``I used AI to express the problem I was experiencing more clearly'' &
    R07, R29\,(4) \\
    \addlinespace[2pt]
    Data documentation for AI systems &
    ``The academic literature on reproducibility keeps pointing to data provenance as a core problem, papers can't be replicated because the exact data pipeline isn't documented. And now with the EU AI Act requiring data documentation for high-risk AI systems (Article 10), this is becoming a regulatory requirement too, not just good practice.'' &
    R36\,(3), R37, R44 \\
    \bottomrule
  \end{tabularx}
\end{table*}

Each entry in the last column of Table~\ref{tab:PracticesThemes} refers to a post, and the number in parentheses gives how many of its excerpts were coded in that category. A single excerpt may be assigned to more than one category when its content addresses several of them.

\FloatBarrier
\section{RQ2: The LLM skills}
\label{sec:rq2}

Table~\ref{tab:SkillsCode} lists the \nSkillCatsWord{} categories of skills we identified. We first discuss the
\nNonTechSkillCatsWord{} categories of non-technical skills, a classification adapted from \citet{rabelo2022nontechnical}, and then the \nTechSkillCatsWord{} categories of technical skills.

\begin{table*}[H]
  \centering
  \scriptsize
  \setlength{\tabcolsep}{4pt}
  \renewcommand{\arraystretch}{1.02}
  \caption{Categories of skills (RQ2). \textit{Excerpts/Posts} indicates the number of coded excerpts (comments or posts) and the number of distinct posts (threads) they come from, respectively.}
  \label{tab:SkillsCode}
  \begin{tabularx}{\textwidth}{
    @{}
    >{\raggedright\arraybackslash}p{2.5cm}
    >{\centering\arraybackslash}p{1.8cm}
    >{\raggedright\arraybackslash}X
    @{}
  }
    \toprule
    \textbf{Category (\nSkillCats{} categories)} &
    \textbf{Excerpts/Posts} &
    \textbf{Consolidated codes} \\
    \midrule
    \multicolumn{3}{@{}l}{\textit{Non-Technical Skills: 4 categories}} \\
    \addlinespace[2pt]
    Critical thinking skills &
    61/17 &
    Understanding how LLMs work; reviewing and verifying AI-generated sources; exercising autonomy and critical thinking when assessing gaps; using AI consciously; distinguishing boilerplate from critical code; identifying when AI goes out of scope; maintaining independence from LLMs. \\
    \addlinespace[2pt]
    Autonomous learning skills &
    27/16 &
    Independent learning; practicing writing and reviewing outputs; reading and interpreting error messages; being able to evaluate code, generate clean code, and understand the fundamentals of AI error correction. \\
    \addlinespace[2pt]
    Problem-solving skills &
    17/11 &
    Understanding how LLMs work; AI literacy and hallucination detection; human critical thinking to discern and identify outdated or generic content; critical and adversarial prompt engineering; maintaining independence and capabilities independent of AI; distinguishing competence from prompting ability. \\
    \addlinespace[2pt]
    Ethics and responsibility skills &
    17/10 &
    Conscious use of AI; knowledge of institutional policy; evaluating data usage responsibly, ethically, and critically; interpreting and tracing generated code; data privacy; developing technical and critical depth beyond prompting. \\
    \midrule
    \multicolumn{3}{@{}l}{\textit{Technical Skills: 5 categories}} \\
    \addlinespace[2pt]
    Programming skills &
    122/20 &
    Grounding in open science, reproducibility, and code versioning; building artifacts before collaborating; developing the practice of coding from scratch, starting with Python, and maintaining clean code; learning the fundamentals to avoid reliance on AI; ability to compare AI-generated code with manually created code and use well-formulated prompts; distinguishing boilerplate from critical code; identifying when AI goes out of scope. \\
    \addlinespace[2pt]
    Computational fundamentals skills &
    38/19 &
    Domain learning for AI/ML application; AI system architecture; combining domain knowledge with technical expertise; mastery of mathematical and statistical fundamentals; a solid foundation in computing concepts, including data structures, algorithms, and logic programming, among others. \\
    \addlinespace[2pt]
    Research skills &
    32/18 &
    Critical analysis; AI literacy, hallucination detection, and ethical/critical evaluation of papers; mapping the genealogy of concepts and using embeddings as a research tool; distinguishing competence from prompting capability; avoiding the delegation of tasks that still need to be learned; practicing the writing and revision of outputs. \\
    \addlinespace[2pt]
    Managing LLM context skills &
    25/12 &
    Calibrating trust in AI output; accountability regarding improper delegation; protecting sensitive data; critical and adversarial prompting; choosing the right tool for each task; maintaining independence and capabilities without AI; recognizing AI's limitations; trusting one's own ability to synthesize information; identifying when AI goes out of scope; and formulating clear, structured requests. \\
    \addlinespace[2pt]
    Data sharing and scientific reproducibility skills &
    21/8 &
    Grounded in open science, reproducibility, and code versioning; responsibly evaluating data usage, interpreting and tracing generated code, and verifying results before reporting. \\
    \bottomrule
  \end{tabularx}
\end{table*}
\FloatBarrier

\subsection{Non-Technical Skills}

\paragraph{Critical thinking skills}
\label{sec:criticalthinking}
This category, adapted from \citet{lee2023}, covers aspects such as creativity, confidence, and reflection. Critical thinking is expected in both academia and industry; with AI, it takes the form of a healthy skepticism that leads to cross-checking results and facts.
Regarding confidence, a commenter in R02 observes: \href{https://www.reddit.com/r/academia/comments/1oz70h8/comment/np9soan/?utm_source=share&utm_medium=web3x&utm_name=web3xcss&utm_term=1&utm_content=share_button}{``\emph{There's indications of more confidence than actual knowledge happening with the use of these tools}''}.
The same thread raises a longer-term concern:
\href{https://www.reddit.com/r/academia/comments/1oz70h8/comment/np9ut1z/?utm_source=share&utm_medium=web3x&utm_name=web3xcss&utm_term=1&utm_content=share_button}{``\emph{use of these tools will stunt people's capacity to develop how to think algorithmically and abstractly. That's where the true skill is}''~(R02)}.

Commenters who use AI heavily in research also stress that judgment remains with the human:
\href{https://www.reddit.com/r/academia/comments/1rc6bt0/using_claude_code_an_ai_coding_agent_has/}{``\emph{To be clear, the human role is still critical. Claude is not good enough to autonomously execute a whole research project without extensive human involvement in ideation, supervision, verification, and auditing. It sometimes makes errors if your instructions are not thorough enough}''~(R03)}.
For others, the problem lies in how AI is used, not in AI itself: \href{https://www.reddit.com/r/academia/comments/1ox9s93/comment/novrojl/?utm_source=share&utm_medium=web3x&utm_name=web3xcss&utm_term=1&utm_content=share_button}{``\emph{It is not AI we should be turning on, it is the behavior of using AI uncritically, unethically that makes people lose critical thinking}''~(R06)}, and a starting point is \href{https://www.reddit.com/r/academia/comments/1ox9s93/comment/novrgpa/?utm_source=share&utm_medium=web3x&utm_name=web3xcss&utm_term=1&utm_content=share_button}{``\emph{Critiquing, enhancing, and augmenting options presented by AI}''~(R06)}. Some go further and argue that humans should \href{https://www.reddit.com/r/PhD/comments/1kdstvd/comment/mqjy9sj/?utm_source=share&utm_medium=web3x&utm_name=web3xcss&utm_term=1&utm_content=share_button}{``\emph{first move beyond critical thinking to the ability to see the bigger picture, recognize patterns, and understand how different elements interact within a complex system}''~(R29)}. The recommendation, in short, is not to stop using AI, but to reflect on and interpret what it produces, and to delegate simple tasks~\href{https://www.reddit.com/r/academia/comments/1ox9s93/comment/novra7q/?utm_source=share&utm_medium=web3x&utm_name=web3xcss&utm_term=1&utm_content=share_button}{(R06)}.

Because LLMs hallucinate, commenters see validating outputs as a skill in which academics need training:
\href{https://www.reddit.com/r/academia/s/nPPJbiOffq}{``\emph{Academics may need to become validators of knowledge trained and paid to detect flaws, AI hallucinations, and reproducibility issues focusing on areas that AI alone (or AI cross-validation) cannot validate}''~(R05)}.
Another recommends:
\href{https://www.reddit.com/r/academia/comments/1ox9s93/comment/nq5qrqv/?utm_source=share&utm_medium=web3x&utm_name=web3xcss&utm_term=1&utm_content=share_button}{``\emph{The real solution is: teach students how to verify AI output, require transparency about when AI is used, evaluate understanding, keep the human in the loop}''~(R06)}. Similar advice appears in R21, \href{https://www.reddit.com/r/PhD/comments/1m906sy/comment/n53mnm8/?utm_source=share&utm_medium=web3x&utm_name=web3xcss&utm_term=1&utm_content=share_button}{``\emph{Always check and double check LLM input, there are way too many incidents of false information or baffling hallucinations from LLMs}''}, and in R31~\href{https://www.reddit.com/r/PhD/comments/1rv2z60/comment/oappzul/?utm_source=share&utm_medium=web3x&utm_name=web3xcss&utm_term=1&utm_content=share_button}{(R31)}. Validating, however, requires knowing the subject:
\href{https://www.reddit.com/r/PhD/comments/1kdstvd/comment/mqda7mq/?utm_source=share&utm_medium=web3x&utm_name=web3xcss&utm_term=1&utm_content=share_button}{``\emph{you need to know when to use it and when not to, and ofc, how to interpret its results}''~(R29)}, which creates a circularity noted in R46:
\href{https://www.reddit.com/r/MachineLearning/comments/1btuizd/comment/ls1s10b/?utm_source=share&utm_medium=web3x&utm_name=web3xcss&utm_term=1&utm_content=share_button}{``\emph{As people rely on LLMs to replace the skills that they should be developing, they will have a reduced ability to verify the LLM is producing correct output}''~(R46)}. Critical thinking about AI output thus depends on the very skills that uncritical use of AI may prevent students from developing.

\paragraph{Autonomous learning skills}
We group posts that refer to active learning skills, that is, autonomous student learning without dependence on AI as the main source: \href{https://www.reddit.com/r/PhD/comments/1rzkuab/comment/oft2hnl/?utm_source=share&utm_medium=web3x&utm_name=web3xcss&utm_term=1&utm_content=share_button}{``\emph{[...] don't use AI as a crutch}''~(R25)}. Users highlight the importance of developing metacognitive skills and autonomous learning: \href{https://www.reddit.com/r/PhD/s/9Kpemm3PAd}{``\emph{don't outsource anything you need to learn to AI}''~(R30)}. Because LLMs answer in natural language, they are easy to consult for everything, which can create dependence. As one post puts it, \href{https://www.reddit.com/r/MachineLearning/comments/1s00v5r/p_opensource_ml_homeworks_with_autotests/}{``\emph{students are adults who need to learn to read error messages}''~(R42)}: identifying errors requires knowledge the student must build on their own. Another commenter advises beginners to \href{https://www.reddit.com/r/compsci/comments/ajdvmx/comment/eev50mc/?utm_source=share&utm_medium=web3x&utm_name=web3xcss&utm_term=1&utm_content=share_button}{``\emph{start with baby steps, try to complete some basic tutorials, online courses}''~(R53)}.
The risk of skipping this stage is illustrated by a user who reports that \href{https://www.reddit.com/r/PhD/comments/1s3h53i/comment/ocg1lbi/?utm_source=share&utm_medium=web3x&utm_name=web3xcss&utm_term=1&utm_content=share_button}{``\emph{junior engineers could prompt their way to decent code but completely fell apart when they had to debug something novel}''~(R30)}: using prompts helps generate code quickly, but this does not guarantee real engineering skills, especially if difficult problems arise.
AI accelerates task delivery, but a student who can read the output of an LLM is not necessarily able to reproduce it, or to fix it, without assistance. The same user summarizes this as \href{https://www.reddit.com/r/PhD/comments/1s3h53i/comment/ocg1lbi/?utm_source=share&utm_medium=web3x&utm_name=web3xcss&utm_term=1&utm_content=share_button}{``\emph{you can't outsource the struggle and still get the skill}''~(R30)} and recommends that students \href{https://www.reddit.com/r/MachineLearning/comments/1olehrk/comment/nmhjttw/?utm_source=share&utm_medium=web3x&utm_name=web3xcss&utm_term=1&utm_content=share_button}{``\emph{[...] your top priority skill is learning how to learn [...]}''~(R40)}. In practice, this means limiting how much one copies, as a student in R27 does: \href{https://www.reddit.com/r/PhD/comments/1l7fcfu/comment/mwwhoh6/?utm_source=share&utm_medium=web3x&utm_name=web3xcss&utm_term=1&utm_content=share_button}{``\emph{I try not to copy and paste too much so I fully understand what the code is doing}''~(R27)}.

\paragraph{Problem-solving skills}
Problem solving complements autonomous learning: commenters stress the ability to connect concepts to practical problems and to identify what the problem actually is before involving an LLM. One account illustrates this: \href{https://www.reddit.com/r/MachineLearning/comments/1sj6sas/comment/ofqmdqa/?utm_source=share&utm_medium=web3x&utm_name=web3xcss&utm_term=1&utm_content=share_button}{``\emph{[...] good intuitions to come up with new approaches and real experience to work with an AI agent to identify problem}''~(R34)}. Another commenter describes decomposing the problem before turning to an LLM: \href{https://www.reddit.com/r/MachineLearning/comments/1mln24c/comment/n7urqg6/?utm_source=share&utm_medium=web3x&utm_name=web3xcss&utm_term=1&utm_content=share_button}{``\emph{what is the overall objective -> what are the sub procedures needed to solve said objective -> what are the helper functions and libraries needed to solve each sub problem -> GPT from there}''~(R44)}.

In this view, the valuable skill \href{https://www.reddit.com/r/MachineLearning/comments/1s9gu8i/comment/odo7yvy/?utm_source=share&utm_medium=web3x&utm_name=web3xcss&utm_term=1&utm_content=share_button}{``\emph{[...] is identifying the actual issue and guiding an LLM to vibe-code a fix}''~(R47)}, because \href{https://www.reddit.com/r/compsci/comments/1kc9zyz/comment/mq7rt0l/?utm_source=share&utm_medium=web3x&utm_name=web3xcss&utm_term=1&utm_content=share_button}{``\emph{the most important thing isn't creating code for the sake of creating it, it's using code to solve the problem you set out to solve}''~(R48)}. As a commenter in R20 puts it: \href{https://www.reddit.com/r/PhD/comments/1shs8id/comment/ofm8wc6/?utm_source=share&utm_medium=web3x&utm_name=web3xcss&utm_term=1&utm_content=share_button}{``\emph{The tool automates downloading and parsing, not thinking. You still have to read, choose what to trace, and interpret the result}''}. In R21, another commenter warns that LLMs do not know the researcher's field: \href{https://www.reddit.com/r/PhD/comments/1m906sy/comment/n539t8c/?utm_source=share&utm_medium=web3x&utm_name=web3xcss&utm_term=1&utm_content=share_button}{``\emph{They are, however, just LLMs, they understand language really well, but likely not your field of research (unless it is linguistics or semantics). It doesn't understand your field, and you should be critical about linguistic conclusions it reaches}''}.

Knowing the field beforehand helps the student analyze the problem and write precise prompts. When this knowledge is present, the gains can be large, as one user reports in R03: \href{https://www.reddit.com/r/academia/s/nOdmanQpE6}{``\emph{Previously, it would take a lot of time reading the manual, finding the right drivers, command syntax, and troubleshooting everything sometimes this would take days. Now, I just need to connect the hardware physically and tell Claude Code what I have connected}''}. Even then, using these tools requires \href{https://www.reddit.com/r/academia/comments/1rc6bt0/comment/o92rzvf/?utm_source=share&utm_medium=web3x&utm_name=web3xcss&utm_term=1&utm_content=share_button}{``\emph{constant experimentation, probing, use of controls, checking assumptions, inspection of intermediate outputs, and sanity checks}''~(R03)}.
For activities that require imagination, however, a researcher in R29 advises against using LLMs, because doing that work is what built their understanding and conviction:
\href{https://www.reddit.com/r/PhD/comments/1kdstvd/comment/nw3mc2b/?utm_source=share&utm_medium=web3x&utm_name=web3xcss&utm_term=1&utm_content=share_button}{``\emph{the imagination work (developing hypotheses, experimental methods, and evaluating results in the context of the literature), I would not have developed an understanding of the subject, sufficient confidence in my beliefs or conviction to defend my conclusions, to finish the PhD}''}. 
A researcher must be able to defend their conclusions, and this comment suggests that the conviction needed to do so comes from having done the work.

\paragraph{Ethics and responsibility skills}
\label{sec:Ethics and responsibility skills}
Graduate programs, funding agencies, and institutions increasingly issue guidelines on the ethical and responsible use of AI in research. One comment raises concerns about \href{https://www.reddit.com/r/academia/comments/1ox9s93/comment/novrojl/?utm_source=share&utm_medium=web3x&utm_name=web3xcss&utm_term=1&utm_content=share_button}{``\emph{the behavior of using AI uncritically, unethically that makes people lose critical thinking}''~(R06)}. In the same thread, another commenter describes using AI as a substitute for one's own ideas as \href{https://www.reddit.com/r/academia/comments/1ox9s93/comment/nu6zjf1/?utm_source=share&utm_medium=web3x&utm_name=web3xcss&utm_term=1&utm_content=share_button}{``\emph{it’s outsourcing thinking and calling it efficiency}''~(R06)}.
Unverified and unattributed use also shows in the quality of student work, as an instructor notes in R29: \href{https://www.reddit.com/r/PhD/comments/1kdstvd/comment/mqfzi3z/?utm_source=share&utm_medium=web3x&utm_name=web3xcss&utm_term=1&utm_content=share_button}{``\emph{[...] I've seen the way grad students use AI without attribution even when it's allowed (with attribution) and I don't understand how even they don't notice their responses are just like a handful of others in the class. It's unoriginal and surface level}''}. 

Others point students to institutional rules: \href{https://www.reddit.com/r/PhD/comments/1l7fcfu/comment/mww9ose/?utm_source=share&utm_medium=web3x&utm_name=web3xcss&utm_term=1&utm_content=share_button}{``\emph{What you really should concern yourself with is your university and programs policies on its use as well as how your advisors, committees and collaborators feel about how you are using it}''~(R27)}, and another commenter in the same thread also recommends checking the university's guidelines~\href{https://www.reddit.com/r/PhD/comments/1l7fcfu/comment/mwwt3jg/?utm_source=share&utm_medium=web3x&utm_name=web3xcss&utm_term=1&utm_content=share_button}{(R27)}.

Another issue involves ethical permissions regarding the submission of data for analysis by AI tools, specifically concerning privacy violations~\href{https://www.reddit.com/r/PhD/comments/1kgsuyf/comment/mr1b7fq/?utm_source=share&utm_medium=web3x&utm_name=web3xcss&utm_term=1&utm_content=share_button}{\emph{(R28)}}. Students therefore need to know which data they are authorized to share with these tools, and under which protocols. Some commenters call for shared rules, such as \href{https://www.reddit.com/r/MachineLearning/comments/wiqjxv/comment/ije9unj/?utm_source=share&utm_medium=web3x&utm_name=web3xcss&utm_term=1&utm_content=share_button}{``\emph{One of the solution is to focus on AI ethics and an universal agreement and adaptation of a set of rules on how and where to use ML in its finest form}''~(R32)}, and note that regulation is already arriving, with the \href{https://www.reddit.com/r/MachineLearning/s/VWfCFiJPj0}{``\emph{EU AI Act requiring data documentation for high-risk AI systems (Article 10)}''~(R36)}. Ethical use, in these accounts, requires students to keep up with rules that are still changing.

\subsection{Technical Skills}
\paragraph{Programming skills}
This category groups comments about the programming skills students need and how to develop them. Commenters argue that students should build a solid technical and conceptual foundation, beyond a superficial mastery of the tools, because this foundation is what allows them to understand, reproduce, and critically evaluate solutions: \href{https://www.reddit.com/r/MachineLearning/comments/1sdmn97/comment/oekd6m7/?utm_source=share&utm_medium=web3x&utm_name=web3xcss&utm_term=1&utm_content=share_button}{``\emph{gaps are precisely where true skill lies}''~(R33)}. There are references to the technical skills frequently required in the programming field, including knowledge of data structures and algorithms, data engineering tools, cloud platforms, and mathematical fundamentals such as statistics and linear algebra.

Similarly, it is recommended that students write code for long enough to experience the consequences of their own decisions, because \href{https://www.reddit.com/r/MachineLearning/comments/1olehrk/comment/nmhjttw/?utm_source=share&utm_medium=web3x&utm_name=web3xcss&utm_term=1&utm_content=share_button}{``\emph{the shorter the feedback loop, the faster you learn}''~(R40)}.
Moreover, one user emphasizes that \href{https://www.reddit.com/r/MachineLearning/comments/1sdmn97/comment/oejwbdp/?utm_source=share&utm_medium=web3x&utm_name=web3xcss&utm_term=1&utm_content=share_button}{``\emph{if you don't fundamentally understand what you're programming at a basic level and the code itself to verify if it's doing what the method should do, you'll almost certainly make mistakes}''~(R33)}.
The same commenter adds that \href{https://www.reddit.com/r/MachineLearning/comments/1sdmn97/comment/oejwbdp/?utm_source=share&utm_medium=web3x&utm_name=web3xcss&utm_term=1&utm_content=share_button}{``\emph{you still need to develop your skills to know when the generated code is correct or not}''~(R33)}, and a commenter in R27 stresses the importance of being able to code without LLMs~\href{https://www.reddit.com/r/PhD/comments/1l7fcfu/comment/mwwkstc/?utm_source=share&utm_medium=web3x&utm_name=web3xcss&utm_term=1&utm_content=share_button}{(R27)}. This matters because \href{https://www.reddit.com/r/PhD/comments/1kdstvd/comment/mqechvc/?utm_source=share&utm_medium=web3x&utm_name=web3xcss&utm_term=1&utm_content=share_button}{``\emph{programming is basically universally relevant for any STEM lab as a required skill}''~(R29)}, and because, as another commenter warns, \href{https://www.reddit.com/r/MachineLearning/comments/1s9gu8i/comment/odq36bg/?utm_source=share&utm_medium=web3x&utm_name=web3xcss&utm_term=1&utm_content=share_button}{``\emph{you have to use tools in an intelligent way to enhance your abilities, and not just offload your capabilities}''~(R47)}. Commenters also list the errors that make this ability necessary, such as calls to non-existent functions and unnecessary code (see Section~\ref{sec:context}).

Generating code with AI is thus not the problem; being unable to debug something new is. A commenter in R30 summarizes this as
\href{https://www.reddit.com/r/PhD/comments/1s3h53i/comment/ocg1lbi/?utm_source=share&utm_medium=web3x&utm_name=web3xcss&utm_term=1&utm_content=share_button}{``\emph{You cant outsource the struggle and still get the skill}''~(R30)}, and another, in R33, proposes a division of labor: \href{https://www.reddit.com/r/MachineLearning/s/PAsk9e9kv4}{``\emph{use LLMs to write the boring parts of the code and write the core parts themselves}''~(R33)}.

\paragraph{Computational fundamentals skills}
This category groups comments about the foundational knowledge in computing, mathematics, and statistics that students need, regardless of the tools they use.

In various accounts, commenters identify topics for building this foundation: \href{https://www.reddit.com/r/PhD/comments/1grccq5/comment/lx520bn/?utm_source=share&utm_medium=web3x&utm_name=web3xcss&utm_term=1&utm_content=share_button}{``\emph{combine code and statistics, data science is your field}''~(R26)} and \href{https://www.reddit.com/r/PhD/comments/1grccq5/comment/lx5590m/?utm_source=share&utm_medium=web3x&utm_name=web3xcss&utm_term=1&utm_content=share_button}{``\emph{Python and matlab are your best bets for getting started. Learn basic data structures and algorithms, and learn how to code professionally}''~(R26)}.

Others emphasize that practice and theory must go hand in hand: \href{https://www.reddit.com/r/MachineLearning/comments/1olehrk/comment/nmhjwyn/?utm_source=share&utm_medium=web3x&utm_name=web3xcss&utm_term=1&utm_content=share_button}{``\emph{Take internships in software positions so you learn design. Don't focus too much on the algorithms/data structures, read more modern dev books}''~(R40)}; another commenter in the same thread recommends focusing on good practices in software engineering, coding, algorithms, data structures, and software design~\href{https://www.reddit.com/r/MachineLearning/comments/1olehrk/comment/nntynjz/?utm_source=share&utm_medium=web3x&utm_name=web3xcss&utm_term=1&utm_content=share_button}{(R40)}. Finally, another concludes that \href{https://www.reddit.com/r/compsci/comments/1kc9zyz/comment/mqe1t1w/?utm_source=share&utm_medium=web3x&utm_name=web3xcss&utm_term=1&utm_content=share_button}{``\emph{languages come and go. Approaches change, but what sets you apart is your ability to solve problems}''~(R48)}. In these accounts, fundamentals are what outlast changes in languages and tools, LLMs included.

\paragraph{Research skills}
This category groups comments about reading, reviewing, and writing research. Commenters stress that reviewing remains a human responsibility: \href{https://www.reddit.com/r/academia/comments/1o0owyq/comment/niav308/?utm_source=share&utm_medium=web3x&utm_name=web3xcss&utm_term=1&utm_content=share_button}{``\emph{Reading and reviewing are also very important, but we should never entrust writing to AI}''~(R05)}, and \href{https://www.reddit.com/r/academia/comments/1qwed3p/comment/o3om32b/?utm_source=share&utm_medium=web3x&utm_name=web3xcss&utm_term=1&utm_content=share_button}{``\emph{researchers need to hone their skills; if they are not critical when making comparisons, synthesizing ideas, and communicating their results, they will not last long as competent researchers}''~(R08)}. One reason is accountability: \href{https://www.reddit.com/r/academia/comments/1qwed3p/comment/o3slxus/?utm_source=share&utm_medium=web3x&utm_name=web3xcss&utm_term=1&utm_content=share_button}{``\emph{LLMs have no credibility and never will because they have no accountability}''~(R08)}. In these accounts, finding and filtering sources to build a coherent argument is itself formative for a new researcher, because it is how they learn to make connections and situate their work.

Research skills also extend to the code that supports research. One commenter recommends documenting code by \href{https://www.reddit.com/r/MachineLearning/comments/1gksoi7/comment/lvo7wox/?utm_source=share&utm_medium=web3x&utm_name=web3xcss&utm_term=1&utm_content=share_button}{``\emph{explaining in the comments why each action is necessary}''~(R41)}, and another argues that \href{https://www.reddit.com/r/PhD/s/hVqrXLvahp}{``\emph{if AI is intended to complement my programming skills, I need radical academic skills}''~(R25)}. In these accounts, research skills are not acquired by delegating sources, code, or literature reviews to AI, but by doing the work of connecting them to a research problem.

\paragraph{Managing LLM context skills}
\label{sec:context}
Another frequently mentioned skill is providing adequate context to LLMs: \href{https://www.reddit.com/r/PhD/comments/1kdstvd/comment/mqng6it/?utm_source=share&utm_medium=web3x&utm_name=web3xcss&utm_term=1&utm_content=share_button}{``\emph{we'll still need people who can ask the right questions in the right ways to make the most of them as tools}''~(R29)}. A commenter in R54 states: \href{https://www.reddit.com/r/PhD/comments/1pgkhul/comment/nssv3zc/?utm_source=share&utm_medium=web3x&utm_name=web3xcss&utm_term=1&utm_content=share_button}{``\emph{I can provide the exact prompt for the code I want}''~(R54)}. Even a good prompt, however, requires reviewing the output: \href{https://www.reddit.com/r/compsci/comments/1kc9zyz/comment/mq6o0rr/?utm_source=share&utm_medium=web3x&utm_name=web3xcss&utm_term=1&utm_content=share_button}{``\emph{I formulate prompts to generate code [...], and the default response might not fit perfectly. But you can steer it toward your style guidelines [...] calls to non-existent functions or slightly incorrect function signatures were the biggest issues I encountered when reviewing AI generated code blocks}''~(R48)}. Providing context and structuring a problem for an LLM thus requires prior knowledge of the task, because only someone who knows the libraries involved can notice a parameter that does not exist.

Commenters also recommend interrogating the model about what it produced: \href{https://www.reddit.com/r/MachineLearning/comments/1sdmn97/comment/oenshju/?utm_source=share&utm_medium=web3x&utm_name=web3xcss&utm_term=1&utm_content=share_button}{``\emph{ask these llms what they generated and why. You can read the code, and write your own changes to it}''~(R33)}. In this view, managing context is not only writing prompts, but steering the conversation, asking for justifications, and taking over when the model's output is wrong, which again depends on the student's own knowledge of the task.

\paragraph{Data sharing and scientific reproducibility skills}
This category groups comments about sharing data and code, and about making research reproducible.

Commenters point out the costs of not sharing: in R10, a user explains that the lack of access to past research results slows down progress, because others cannot replicate those results and build on them~\href{https://www.reddit.com/r/academia/s/LXVDWnRLzM}{(R10)}. In R36, another user notes that some journals already require reproducibility checks~\href{https://www.reddit.com/r/MachineLearning/comments/1r7usv0/comment/o69jv8e/?utm_source=share&utm_medium=web3x&utm_name=web3xcss&utm_term=1&utm_content=share_button}{(R36)}, and that \href{https://www.reddit.com/r/MachineLearning/comments/1r7usv0/comment/o66p5v1/?utm_source=share&utm_medium=web3x&utm_name=web3xcss&utm_term=1&utm_content=share_button}{``\emph{the gap is most apparent in teams with less structured workflows}''~(R36)}. In R44, a commenter goes further, noting that \href{https://www.reddit.com/r/MachineLearning/comments/1mln24c/comment/n85f5wv/?utm_source=share&utm_medium=web3x&utm_name=web3xcss&utm_term=1&utm_content=share_button}{``\emph{failing to make research code or data available can, in some cases, be considered scientific misconduct}''~(R44)}.

In these accounts, sharing data and code is part of doing research, not an addition to it. This skill becomes more important as LLMs make it easier to produce custom code and pipelines that only their author, or no one, fully understands.

Table~\ref{tab:SkillsThemes} maps the excerpts to the \nSkillCatsWord{} categories of skills, \nNonTechSkillCatsWord{} non-technical and \nTechSkillCatsWord{} technical. Posts R07, R11, R12, R14, R17, and R52 appear in the practice categories but in none of the skill categories.
\begin{table*}[!t]
  \centering
  \fontsize{6.6}{7.6}\selectfont
  \renewcommand{\arraystretch}{0.82}
  \caption{Categories of skills (RQ2), with a representative quote and the posts of the coded excerpts. A number in parentheses is the number of coded excerpts from that post.}
  \label{tab:SkillsThemes}
  \begin{tabularx}{\textwidth}{@{}>{\raggedright\arraybackslash}p{2.1cm} X >{\raggedright\arraybackslash}p{3.4cm}@{}}
    \toprule
    \textbf{Skill category (\nSkillCats)} & \textbf{Representative citation} & \textbf{Examples} \\
    \midrule
    Critical thinking skills &
    ``Academics may need to become validators of knowledge---trained and paid to detect flaws, AI hallucinations, and reproducibility issues---focusing on areas that AI alone (or AI cross-validation) cannot validate.'' &
    R01\,(3), R02\,(3), R03, R05, R06\,(9), R21\,(3), R22, R29\,(10), R31, R32, R33\,(2), R45\,(4), R46\,(5), R47\,(6), R48\,(4), R51\,(5), R54\,(2) \\
    \addlinespace[1.5pt]
    Problem-solving skills &
    ``Using these tools aligns well with the nature of STEM science, where the tools---whether software, hardware, or wetware---are inherently critical and may require extensive iterative problem-solving, constant experimentation, investigation, the use of controls, assumption verification, the inspection of outputs, and sanity checks.'' &
    R02, R03\,(2), R20, R21, R29\,(2), R34, R39, R44, R46, R47, R48\,(5) \\
    \addlinespace[1.5pt]
    Autonomous learning skills &
    ``Words of caution, don't use AI as a crutch. I've fought every step of the way to remain `pure' I this sense. I think not using AI would be a bad thing, but you can't just have it mass-produce assignments for you. Plagiarism is a real thing and you also want to learn the subjects'' &
    R13, R16, R25, R27\,(4), R30, R31, R33\,(2), R39, R40, R42, R43, R44\,(3), R46, R48\,(3), R53\,(2), R54\,(3) \\
    \addlinespace[1.5pt]
    Ethics and responsibility skills &
    ``What you really should concern yourself with is your university and programs policies on its use as well as how your advisors, committees and collaborators feel about how you are using it provided it is permissible or not forbidden by your university or program policies.'' &
    R01, R06\,(2), R25, R27\,(5), R28, R29\,(2), R32, R36\,(2), R37, R46 \\
    \addlinespace[1.5pt]
    Managing LLM context skills &
    ``Since I can precisely prompt what code I want from it, I can get the exact code I want most of the time'' &
    R03\,(2), R06\,(3), R21\,(2), R29, R33\,(5), R44, R45\,(2), R46, R47\,(2), R48\,(3), R50, R54\,(2) \\
    \addlinespace[1.5pt]
    Research skills &
    ``What if doctoral training included structured review practice, AI literacy, and AI ethics for reviewing?'' &
    R02, R04\,(2), R05\,(2), R06\,(2), R08\,(5), R10\,(3), R18, R20, R21\,(2), R22, R25\,(2), R29\,(4), R30, R31, R33, R39, R41, R44 \\
    \addlinespace[1.5pt]
    Data sharing and scientific reproducibility skills &
    ``You already know which charts you want to create, how your data needs to be located and structured to generate a chart, and have a clear idea of how to handle outliers (missing data, imbalanced data, etc.). These skills are what make research both challenging and valuable.'' &
    R10\,(6), R28\,(3), R29, R36\,(4), R38, R44\,(4), R47, R54 \\
    \addlinespace[1.5pt]
    Programming skills &
    ``I would treat that script as a prototype and then rebuild pieces yourself until you can explain what every line does. Take a basic ML course, do small coding exercises, and use AI as a helper to debug and clarify, not as the main author'' &
    R02, R03\,(2), R13\,(2), R21\,(2), R25, R26\,(2), R27\,(4), R29\,(3), R30, R32, R33\,(32), R35, R39, R40\,(2), R41\,(9), R44\,(17), R47\,(7), R48\,(16), R51\,(3), R54\,(15) \\
    \addlinespace[1.5pt]
    Computational fundamentals skills &
    ``There was a time when many in academia would absolutely refuse to program\ldots{} they'd call it a fad and opt for pen and paper approaches. Now, programming is basically universally relevant for any STEM lab as a required skill'' &
    R09, R14, R15, R19, R23, R24, R25, R26\,(5), R29, R38\,(2), R39, R40\,(5), R41, R43, R44\,(6), R47, R48\,(2), R49\,(4), R53\,(2) \\
    \bottomrule
  \end{tabularx}
\end{table*}
\FloatBarrier
\section{Discussion}
\label{sec:discussion}

Our findings suggest that LLMs are no longer peripheral tools in graduate research.
Commenters on Reddit report using them in several research-related
activities, including writing, programming, debugging, testing, documentation,
literature exploration, and experiment iteration. These practices suggest that
LLMs are becoming part of the everyday infrastructure of research work, not
only as writing assistants, but also as coding partners, search tools, and
agents that support repetitive or technical tasks.
At the same time, the posts reveal an important tension. LLMs can accelerate
research work, but they also make it easier for students to delegate tasks that
are central to their own formation as researchers. In the accounts we analyzed, the main issue is not the use
of LLMs itself, but the lack of validation, judgment, and accountability when
students accept generated outputs without understanding them. In RQ1 and
RQ2, we observed that productive use depends on combining traditional skills,
such as programming, debugging, research design, scientific writing, and
critical thinking, with LLM-specific skills, such as prompting, context
management, hallucination detection, and output verification.

\paragraph{Comparison with existing frameworks}
Several of our categories correspond to competencies proposed by \citet{annapureddy2025}: their ability to assess the output of generative AI tools corresponds to our critical thinking skills; skill in prompting, to managing LLM context; ability to program and fine-tune generative models, partially to programming skills; knowledge of ethical implications, to ethics and responsibility skills; and ability to continuously learn, to autonomous learning. Other categories are specific to research and do not appear in their framework, such as data sharing and scientific reproducibility, research skills, and the use of LLMs for accessibility. On the other hand, competencies such as the detection of AI-generated content and knowledge regarding legal aspects appear only infrequently in our corpus. This comparison suggests that general AI literacy frameworks are a starting point, but that research training needs competencies tied to the integrity of research outputs.

\paragraph{Delegation, not prohibition}
An implication of these findings is that graduate education should avoid
defining LLM use as a binary choice between prohibition and unrestricted
adoption. The accounts suggest a more nuanced position: students need to learn which
parts of research work can be delegated, which parts can be supported, and which
parts must remain under human responsibility. For example, using an LLM to
generate boilerplate code, refactor a function, draft documentation, or suggest
tests may be appropriate when the student can inspect and validate the result.
In contrast, delegating theoretical reasoning, methodological decisions,
citation validation, or interpretation of results without scrutiny creates risks
for both learning and research quality.

\paragraph{Programming in empirical research}
A second implication concerns the changing nature of programming in empirical
research. Several posts indicate that students and researchers use
LLMs to write scripts, refactor code, generate tests, explain errors, and build
small tools. However, these same posts emphasize that researchers still need to
understand the code they run. This is particularly important in empirical
software engineering and machine learning, where errors in scripts, pipelines,
or experimental configurations may silently compromise results. In this context,
the researcher becomes less of a passive code consumer and more of a reviewer,
tester, and maintainer of AI-generated artifacts.

\paragraph{Implications for graduate programs}
If students use LLMs mainly to avoid difficulty, they can complete tasks without developing the
skills that those tasks were meant to teach. If LLMs are used as tools
for explanation, feedback, comparison, and controlled automation, they may help
students engage more deeply with research work. The central challenge for
graduate programs is therefore not to decide whether students should use LLMs,
but to define what responsible use looks like in research training. Based on our categories, we suggest that programs:
\begin{itemize}
\item make verification an explicit learning activity, for instance by asking students to check every reference and every generated analysis they include in their work, and to report how they did it;
\item assess whether students can explain what they deliver, such as code walkthroughs or oral examinations, while offering accommodations to students who rely on LLMs for accessibility;
\item discuss with students, early in their training, which research tasks they should not delegate while they are still learning them, such as literature synthesis, study design, and interpretation of results;
\item teach data documentation and reproducibility practices as part of research methods courses, since LLMs make it easier to produce pipelines that no one fully understands.
\end{itemize}

\subsection{Limitations}
\label{sec:limitations}

This study has some limitations. First, our findings are based exclusively on
Reddit posts, which represent self-reported experiences shared by users of a
specific online platform. Reddit users may not be representative of the broader
population of graduate students, and discussions are naturally biased toward
individuals willing to publicly share their concerns, frustrations, and
experiences. In addition, the anonymity of Reddit makes it impossible to
verify users' academic status, research experience, or actual use of LLMs, and our corpus includes not only students, but also advisors, instructors, and industry professionals commenting on students.
Therefore, our results should be interpreted as perceptions and reported
practices rather than direct observations of student behavior.

Second, the corpus is concentrated in a few threads. Four posts (R33, R44, R48, and R54) account for 80 of the excerpts coded as programming skills, and some categories, such as Data documentation for
AI systems and accessibility, come from four or fewer posts. To mitigate this, we report the number of distinct posts per category, and we do not interpret the number of excerpts as a measure of prevalence. \nPreLLMPostsWordCap{} included posts (R09, R10, R38, R43, R49, and R53) were published before 2022 and therefore before the popularization of LLM-based chat assistants. Although published before 2022, these posts address discussions regarding studies and programming involving AI (for instance, R49, published in 2014) and also cover research into ML, as well as the practices, methods, and competencies that underpin the current use of these tools.

Third, the screening and qualitative coding involve subjective judgment. The first author screened all retrieved posts, a second author reviewed the resulting set, and we complemented this process with an LLM. Different search queries, subreddits, or qualitative coding decisions could have led to partially different categories.

Regarding language, we limited the corpus to English and Portuguese. Portuguese was included because it is the native language of the authors, and English because most discussions on Reddit, including those in the selected communities, are in English. Our analysis revealed that all posts with an ``included'' status are in English; these posts are available in the supplementary material.

Finally, online discussions tend to emphasize problematic, controversial, or exceptional experiences, potentially underrepresenting successful and routine uses of LLMs in graduate research. Future work could complement our findings through interviews, surveys, or observational studies involving graduate students and advisors from different institutions and research disciplines.

\section{Conclusion}
\label{sec:conclusion}

This study investigated the use of LLMs in graduate research through a GLR based on Reddit, aiming to identify the practices students adopt (RQ1) and the skills they need (RQ2). We collected \nRetrieved{} posts from \nSubredditsWord{} research-related subreddits, of which \nIncluded{} were included for qualitative analysis. Commenters describe LLMs as boosting productivity by providing quick solutions, but also warn that delegating the effortful parts of research work without verification may prevent students from developing research skills.

\begin{itemize}
\item In RQ1, we identified \nPracticeCatsWord{} categories of practices: graduate students use LLMs
throughout the research process, particularly for programming, writing, learning, testing,
data analysis, and documentation, with varying levels of
validation.

\item In RQ2, we identified \nSkillCatsWord{} categories of skills: the accounts indicate that students need both traditional
skills, such as programming and critical thinking, and LLM-specific skills, such
as prompting, context management, and output verification, and that the former are what make the latter possible.
\end{itemize}

\subsection{Future work}

As future work, we plan to: (i) investigate whether intensive LLM usage changes how research skills are acquired and retained over time, (ii) develop
evidence-based pedagogical approaches for teaching empirical research in the
presence of AI assistants and agents, and (iii) establish a research agenda for defining, measuring, and assessing researcher skills in the LLM era.

\section*{Declaration of competing interest}
The authors declare that they have no known competing financial interests or personal relationships that could have appeared to influence the work reported in this paper.

\section*{CRediT authorship contribution statement}
Enne Rebeca Silva de Freitas contributed to the study's conceptualization, methodology, literature search, study selection, data extraction, thematic analysis, and drafting of the original manuscript. Breno Alves contributed to the methodology, reviewed the manuscript, and supported the conceptualization of the thematic analysis. Gustavo Pinto and Danilo Ribeiro contributed to the study's conceptualization, methodology, study selection, thematic analysis, validation of the results, and supervision. They also contributed to the review and editing of the manuscript. All authors discussed the results, contributed to the interpretation of the findings, and read and approved the final manuscript.

\section*{Ethics approval and consent}
This study did not involve human participants. It analyzed only publicly available posts and comments and involved no interaction with users (see Section~\ref{sec:ethics}).

\section*{Data availability}
The data supporting this study are available at Zenodo \citep{zenodo2026}. The repository provides the search and qualitative coding materials, including the queries, the scripts, and the post-level mapping between posts, practices, skills, research questions, and proposed categories.

\section*{Declaration of generative AI and AI-assisted technologies in the writing process}
During the preparation of this work the authors used ChatGPT (OpenAI, GPT-6) to assist in translating the manuscript from Portuguese to English and in categorizing the publications according to codes defined by the authors, by comparing the model's decisions with our own, as described in Section~\ref{sec:synthesis}. These tools were not used to define research questions or make decisions regarding the interpretation of the results. After using this tool, the authors reviewed and edited the content as needed and take full responsibility for the content of the published article.

\bibliographystyle{cas-model2-names}

\bibliography{references}

\end{document}